\documentclass[preprint,12pt]{elsarticle}

\usepackage{amssymb}
\usepackage{amssymb,amsthm,amsmath}
\usepackage{multirow}
\usepackage[latin1]{inputenc}
\usepackage{graphicx}
\usepackage{algorithm}
\usepackage{algorithmic}
\usepackage{url}
\newcommand{\R}{\mathbb{R}}

\newcommand{\be}{\begin{equation}}
\newcommand{\ee}{\end{equation}}
\newcommand{\bea}{\begin{eqnarray}}
\newcommand{\eea}{\end{eqnarray}}

\usepackage{tikz}
\definecolor{gold}{rgb}{0.85,.66,0}
\definecolor{cian}{rgb}{.02,.7,.95}
\definecolor{ppp}{rgb}{.7,.3,.82}

\newcommand{\tp} {^{\intercal}}

\usepackage[normalem]{ulem}

\renewcommand{\a}{\mathbf{a}}

\newcommand{\p}{\mathbf{p}}

\renewcommand{\u}{\mathbf{u}}

\renewcommand{\H}{\mathbf{H}}
\newcommand{\I}{\mathbf{I}}

\renewcommand{\d}{\mathbf{d}}
\renewcommand{\P}{\mathbf{P}}

\newcommand{\1}{\mathbf{1}}

\newcommand{\bm}[1]{\mbox{\boldmath{$#1$}}}

\usepackage{subfig}
\graphicspath{{figuras/}{fig_site/}}

\journal{ArXiv  \qquad arXiv.org \qquad Cornell University}

\begin{document}

\begin{frontmatter}



\title{Hopfield Learning-based and Nonlinear Programming methods for Resource Allocation in OCDMA Networks}

\author[a1]{Cristiane A. Pendeza Martinez}
\ead{crismartinez@utfpr.edu.br}
\author[a2]{Taufik Abrão\corref{cor1}}
\ead{taufik@uel.br}
\author[a1]{Fábio Renan Durand}
\ead{fabiodurand@utfpr.edu.br}
\author[a1]{Alessandro Goedtel}
\ead{agoedtel@utfpr.edu.br}

\address[a1]{Universidade Tecnológica Federal do Paraná -- Campus Cornélio Procópio
Avenida Alberto Carazzai, 1640 - CEP 86300-000
Cornélio Procópio - PR - Brasil}
\address[a2]{Department of Electrical Engineering, Londrina State University, Rod. Celso Garcia Cid - PR445,  Po.Box 10.011, CEP: 86057-970, Londrina, PR, Brazil}

\begin{abstract}
This paper proposes the deployment of the Hopfield's artificial neural network (H-NN) approach to optimally assign power in optical code division multiple access (OCDMA) systems. Figures of merit such as feasibility of solutions and complexity are compared with the classical power allocation methods found in the literature, such as Sequential Quadratic Programming (SQP) and Augmented Lagrangian Method (ALM). The analyzed methods are used to solve constrained nonlinear optimization problems in the context of resource allocation for optical networks, specially to deal with the energy efficiency (EE) in OCDMA networks. The promising performance-complexity tradeoff of the modified H-NN is demonstrated through numerical results performed in comparison with classic methods for general problems in nonlinear programming. The evaluation is carried out considering challenging OCDMA networks in which different levels of QoS were considered for large numbers of optical users.
\end{abstract}




\end{frontmatter}


\section{Introduction}
Due to the technological advancement of state-of-the-art media, the research is focused on finding systems with greater efficiency, higher transmission capacities and greater range with fewer repeaters. Nowadays, with the evolution of photonic technology, optical transmission media have become the most feasible option for large-scale information transmission in a fast and reliable way and reaching high transmission rates in several systems.

A fiber optic connection has low loss between its transmitter and receiver, in addition to being able to transmit analog or digital signals. The original signal is converted from electrical to optical through a media converter, in a point-to-point transmission or an optical line terminal (OLT) if  fiber to the home (FTTH) is deployed.

Based on the evolution of the code division multiple access (CDMA) technique in wireless systems, the respective optical technique (OCDMA) was introduced in the mid 1980s \cite{1}, \cite{01}. OCDMA systems were developed using the spectral scattering technique, where each user is indicated by a unique code. Due to the tremendous growth of OCDMA network, several configurations have been established which can be classified into non-coherent and coherent systems \cite{001}. Moreover, OCDMA technology has attracted research interest because of its many advantages, such as asynchronous operation, network flexibility, protocol transparency, simplified control and also making the network potentially safer \cite{2}, \cite{45}.

The resources allocation, such as power and bandwidth, optical networks are decisive in order to make the most efficient as possible the deployment of the available bandwidth in this network mode, as well as, observing that the transmission occurs aiming at minimizing the non-linear effect impacts caused by the physical and structural limitations of the optical fibers during transmission. For this, a robust system is necessary, using encoders and decoders that can perform the transmission safely and efficiently. In this context, a paramount resource allocation problem in OCDMA networks is the power allocation that minimizes the multiple access interference. In solving such optimization problem, the optical network is able to accommodate the largest number of users sharing the same spectrum network, guaranteeing for each user its minimum quality of service (QoS) in terms of minimum data rate and  signal to interference plus noise ratio (SINR) \cite{5}. On the other hand, the power allocation in optical systems must deal with the optical power budget in the sense that it is paramount guarantee the mitigation of the non-linear optical effects due to physical fiber imperfections and construction network limitations.

Works as \cite{8} and \cite{10} for CDMA network and for OCDMA network \cite{32} use an analytical and interactive approach, namely the Verhuslt population model to obtain a new distributed power control algorithm (DPCA). The Verhuslt model has demonstrated high speed of convergence, quality of solutions (proximity of optimal value after convergence), robustness to estimation errors, among others as positive aspects for implementation.

The power control (PC) in OCDMA optical networks appears as a nonlinear optimization problem. In this case, the objective is to establish power levels so that the signal-to-noise ratio plus each user's interference reaches a threshold required for acceptable performance (maximum tolerable error rate) and quality of service required (minimum QoS).
Several engineering problems are modeled as optimization problems which require methods that provide realistic and processed solutions to digital computers quickly. In this way, traditional methodologies such as programming methods can provide inaccurate solutions. Therefore, the search for solutions in such cases becomes important and essential, as the structure of certain problems is complex or there are a large number of possible solutions \cite{24}.

The artificial intelligence (AI) area proposes several techniques and resources in the development of intelligent programs, that is, programs capable of making a decision similar to human \cite{02}. The area of artificial intelligence began to develop in the sense of modeling the brain through the creation of artificial neural networks (ANN), which have the same cognitive and associative properties of the human brain. Problems of difficult treatment in conventional computation, can be approached by using ANN, so that it elaborates effective solutions \cite{24}. In recent years the theory of ANN has made significant progress that has led to the development of more effective tools in solving problems across different areas of knowledge. A condition for these advances is the more efficient use of the available computational resources, which generates an increase in the manipulation capacity of the information.

Since its inception,  ANN deployment has been motivated by the recognition that the human brain computes in an entirely different way from the conventional digital computer highly complex, nonlinear, and parallel computer (information-processing system) \cite{38}, \cite{47} and \cite{44}. Specifically, the ANN procedures deployed to solve nonlinear optimization problems have been developed using penalty parameters \cite{25}, \cite{26} and \cite{28}. The equilibrium points of these networks, corresponding to the solutions of the problem, are obtained through the appropriate choice of penalty parameters that must be sufficiently large to guarantee the convergence of the network. Thus, the choice of a specific ANN structure and it respective parameter values is a complex task performed through empirical techniques, which require a very excessive computational effort and exhaustive {training} steps. In addition, the quality of the final solution also depends on the  parameters fitting \cite{29}.

Some other difficulties related to the convergence process for the network equilibrium points, which represents the solutions of the optimization problem also should be considered. Numerical results presented in \cite{25}, \cite{26} and \cite{28} discuss the infeasible solutions. In this context, the authors of \cite{24} seek an alternative to improve the efficiency of computer simulations and to provide a new methodology for mapping constrained nonlinear optimization problems using the modified Hopfield neural networks (mH-NN). The main characteristics of such methodology deploying {mH-NN} are: (i) no weighting constants; (ii) all structural constraints involved in the constrained nonlinear optimization problem are grouped into a single constraint term; (iii) there is no interference between the optimization term and the restriction term; (iv) no initialization parameter is required for simulation execution.

In this work, an analysis of the dynamic behavior of the optical CDMA network is carried out, while the convergence process towards the optimal power solutions is careful analyzed. By analyzing the numerical and theoretical results, in this work we seek generalizations for the power allocation problem in OCDMA networks aiming at analyzing the applicability of Hopfield-based ANN as a procedure to solve general classes of resource allocation problem. Herein, the analytical solutions are explored through continuous optimization methods, namely sequential quadratic programming (SQP) and augmented Lagrangian method (ALM). It is worth noting that in the literature, the optimization approach commonly applied to solve power assignment problems in optical networks has been heuristic methods, for instance \cite{32}, {\cite{46}.

Based on the formulation of the optimization problem, it is notable that the minimum power allocation problem presents only one point that satisfies Karush-Kuhn-Tucker (KKT) conditions. Thus, in this work, the  nonlinear programming (NLP) classical methods, such as SQP and ALM, were deployed as a baseline to solve the optical power allocation problem.

\noindent {\bf Contribution}: The contribution of this work is threefold. It consists of {\bf a}) propose the use of modified Hopfield-based NN (mH-NN) especially constructed to solve the minimum power allocation problem in OCDMA with QoS guarantee;
{\bf b}) a systematic analysis of the minimum power allocation problem in OCDMA networks using the mH-NN; and {\bf c}) comparing it with the classical optimization methods SQP and ALM which are analytical methods of optimization applied specifically to  solve the minimum optical power allocation problem taking into account the convergence speed, feasibility, complexity and optimality, in implementing realistic OCDMA networks.

The remainder of this document is divided as follows: Section \ref{sec:def} discusses the problem of minimizing power allocation in OCDMA systems based on QoS. The optimization methods used in this work are presented in Section \ref{sec:met}. Also, aspects related to the implementation of the optimization methods applied to the problem of minimum power allocation and numerical results are analyzed in Section \ref{sec:resu}. Finally the Section \ref{sec:Concl} presents the final remarks.

\section{Definition of Power Allocation Problem}\label{sec:def}
The PC is applied to systems users interfere with each other. Performing power control should adjust the transmitted power of all users so that each user's noise ratio plus interference (SNIR) meets a certain threshold required for acceptable performance.

In order to achieve a specific QoS, which is associated with a bit error rate (BER) that is tolerated by the $i$th optical node, the relation carrier-noise interference (CIR) required in the network receiver decoder can be defined as \cite {7}:
\begin{equation}\label{CIR}
\Gamma_i = \frac{G_{ii}p_i}{\sum_{j=1, j\neq i}^KG_{ij} p_j+\sigma_i^2} \geq \Gamma_i^*
\end{equation}
where $p_i$ is the power of the $i$th node, $K$ is the dimensional value of the column vector of transmitted optical power namely $\p = [p_1, p_2, ...... p_K]\tp$, $\sigma_i^2$ is the power of additive white Gaussian noise (AWGN) inherent to the communication system at the receiving node $i$ and $G_{ij}$ is the fiber attenuation between the $j$th transmitting node and $i$-th receiving node. Therefore, the BER is a related QoS metric as well as the SNIR. The SNIR is associated with the CIR as:
\begin{equation}\label{CIR_SNIR}
\gamma_i=\frac{r_c}{r_i}\Gamma_i, \hspace{0.5 cm} i=1,\ldots,U
\end{equation}
where $r_c$ is the chip rate, $r_i$ is the information rate of the $i$th user, $U$ is the number of users of the system and $\Gamma_i$ is the CIR of $i$th user.

The objective of the minimum power allocation problem is to find the minimum transmission power for each system user while satisfying all QoS requirements. Such QoS requirements are basically summarized to the minimum information transmission rate and the maximum tolerable BER. This can be summarized in SNIR using it as a constraint for QoS guarantee. The PC problem can be mathematically described as:
\begin{align}\label{probe}
\min &\hspace{.3cm} J_1(\textbf{p})= \1\tp \textbf{p} =\sum_{i=1}^{K}p_i \notag\\
{\rm s. \ t.} & \quad  \displaystyle \Gamma_i = \frac{G_{ii}p_i}{\sum_{j=1, j\neq i}^KG_{ij} p_j+\sigma_i^2} \geq \Gamma_i^*\hspace{1cm}\\
 & \quad p_{\min}\leq p_i \leq p_{\max}, \  \   \forall i=1,\cdots, K, \notag\\
 & \quad  p_{\min}>0, \hspace{.03 cm} p_{\max}>0\notag
  \end{align}
where $\textbf{1}^T=[1,\ldots,1]$, $\Gamma_i^*$ is the minimum CIR for the $i$th user reach the desired QoS level, $p_{\min}$ is the minimum optical transmission power and $p_{\max}$ is the maximum optical transmission power. Using matrix notation one can express the inequality as:

\begin{equation}
[\I-{\bm\Gamma}^* \, \H]\, \P\geq \bar{\u},
\end{equation}
where $\I$ is the identity matrix, $\H$ is the normalized interference matrix whose elements are given by:
 \begin{equation}
 {H}_{ij}=\left\{\begin{array}{ll}
0, & i=j; \\
 \frac{G_{ij}}{G_{ii}}, & j\neq i.
 \end{array}
 \right., \hspace{0.5cm}
\bar{u}_i=\frac{\Gamma_i^* \sigma_i^2}{G_{ii}}, \hspace{0.5cm}
\ \ \Lambda^* =\left(
  \begin{array}{ccc}
    \Gamma_1^* & ... & 0 \\
    0 & \Gamma_2^* & ... \\
    0 & 0 & ...\Gamma_k^* \\
  \end{array}
\right).\end{equation}
Substituting inequality by equality, the optimized power vector solution is given by \cite{17} \cite{6}:

\begin{equation}\label{tarhuni}
\p^*=[\I-{\bm\Gamma}^* \H]^{-1}\u.
\end{equation}

The PC problem in OCDMA networks can be classified as a non-linear programming problem and not convex due to the constraint of the imposed CIR. From the literature we can cite authors such as \cite{17} who use ant colony optimization (ACO)  method to solve the problem (\ref{probe}).  However, heuristic methods frequently do not generate promising results regarding performance-complexity tradeoff when compared to deterministic methods. Occasionally, one can also find solutions to the problems that are far from optimal values when using such metaheuristic methods. In addition, the input parameters of the heuristic methods,  must be adjusted by the exhaustive search, which often can impact the results of more detailed analyzes if not adjusted properly.

An alternative to problem solving (\ref{probe}) is the use of analytical approaches \cite{6}. In this approach, the determination of solution has greater computational complexity in relation to the heuristic alternatives, but they are able to guarantee the optimality of the solution.

In this context, this work proposes an alternative solution to the problem (\ref{probe}) with the Hopfield network. The use of ANNs to solve optimization problems was first proposed by Hopfield and Tank \cite{11}. Since then it has been explored the possibility of solving problems of optimization with approaches by neural networks.

In particular authors such as \cite{24} introduce a new methodology for mapping restricted nonlinear optimization problems called modified Hopfield Networks in order to bypass network convergence problems and improve the efficiency of computer simulations.
\section{Approached Optimization Methods}\label{sec:met}
The increase in the demand for transmission rate, related in large part by the continuous growth of Internet traffic, implies the need to increase the flexibility and the capacity of the network \cite{34}. However, the degradation of the SNIR appears as a challenge, since the problem of the near-far effect appears together with the interference by multiple access (MAI). In this way, it becomes necessary to establish an efficient management  of the resources, as example, the optical power control is necessary to overcome this problem, increasing performance and optimizing network utilization. This solution can be achieved by solving the optimization problem posed by eq.(\ref{probe}). In this section we  revisit the main characteristics of the optimization methods discussed in this work to solve this problem.

\subsection{Hopfield Artificial Neural Network (H-NN)}\label{sec:RNA}
ANNs are computational models inspired by the central nervous system of an animal, in particular the brain, capable of performing machine learning as well as pattern recognition. They are usually presented as systems of interconnected neurons, which can compute input values, simulating the behavior of biological neural networks. These models are used to solve various engineering problems such as function approximation, pattern classification and optimization. In this work we use the modified H-NN proposed by \cite{24} in a nonlinear constrained optimization problem.

In 1982, John Hopfield presented a network type different from those based on Perceptron \cite{22}. In this model the network presented recurrent connections and was based on the unsupervised learning with the competition among the neurons. This type of artificial neural network architecture has the following characteristics: ({\it i}) typically dynamic behavior; ({\it ii}) ability to memorize relationships; ({\it iii}) possibility of storing information; ({\it iv}) ease analog implementation.

The deployed Hopfield network \cite{22} has the structure as depicted in Figure \ref{rede}, with a single layer in which all neurons are completely interconnected, i.e. all neurons of the network are connected to all the others and themselves where the outputs feed the inputs.

\begin{figure}[!htbp]
\begin{center}
\includegraphics[width=8cm]{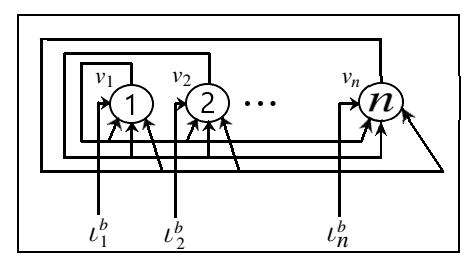}
\caption{Conventional Hopfield Network}\label{rede}
\end{center}
\end{figure}

The simplified couple of expressions governing the continuous-time behavior of each neuron in the Hopfield network are given by:
\begin{equation}\label{1}
\displaystyle
\dot{u}_j(t)=-\eta u_j(t)+\sum_{j=1}^{N} {\bf W}_{ij}v_i(t)+{\bm \iota}_{j}^{b}, \qquad j=1,\ldots,n
\end{equation}

\begin{equation}\label{func_ativacao}
v_j(t)=g(u_j(t))
\end{equation}
where $\dot{u}_j(t)$ is the internal state of the $j$-th neuron, with $\dot{u}_j(t) = du/dt$; $v_j(t)$ is the output of $j$-th neuron; ${\bf W}_{ji}$ is the synaptic weight by $j$th  neuron to $i$th neuron; ${\bm \iota}_{j}^{b}$ is the threshold (bias) applied to $j$th neuron; $g(.)$ is a growing monotonic activation function, which limits the output of the neuron; $\eta u_j(t)$ is a passive decay term.  Observing the expressions (\ref{1}) and (\ref{func_ativacao}) on can verified that the behavior of the Hopfield network is always dynamic, that is, a set of inputs is applied; and then the outputs $v$ are calculated and fed back to the inputs. The output is then recalculated and the process repeats in an iterative manner. Successive iteration sequences produce (decreasingly) changes in network outputs until their values become constant (stable).

Therefore, given any set of initial conditions, can be obtained by second Lyapunov method as presented in \cite{44} a Lyapunov function for the Hopfield network whose neurons are changed one at a time is defined by:
\begin{equation}\label{func_energia1}
{E(t)=-\frac{1}{2}{\bf v}(t)^{T}{\bf W}} {\bf v}(t)-{\bf v}(t)^{T} \bm{\iota}^b
\end{equation}
where the equilibrium points of the network correspond to the values of $v(t)$ that minimize the energy function of the network; ${\bf W}$ is the weight matrix; ${\bm \iota}^b$ is the input vector associated with the power function of the network (\ref{func_energia1}).

From (\ref{func_energia1}) we obtain the expression for its temporal drift, that is:
\begin{equation}\label{func_energia}
\dot{E}(t)=\frac{dE(t)}{dt}=(\bm \nabla_{\bf v}E(t))^T {\bf v}(t)
\end{equation}
where ${\bm \nabla}_{\bf v}$ is the operator gradient in relation to the vector $\bf v$. As long as the weight matrix is symmetric, ${\bf W}={\bf W}^T$, we have:
\begin{equation}\label{gradiente1}
\bm{\nabla}_{\bf v} E(t)=-{\bf W} {\bf v}(t)-{\bm \iota}^b
\end{equation}

From (\ref{1}) assuming that the passive decay term is zero, we conclude the following result with (\ref{gradiente1}):
\begin{equation}\label{gradiente}
{\bm \nabla}_{\bf v} E(t)=-\dot{u}(t)
\end{equation}

Using the above relations we obtain the expression for the derivative of the function $ E(t)$:
\begin{equation}\label{func_energia}
\dot{E(t)}=-\displaystyle \sum_{j=1}^{n}\underbrace{(\dot{u}_j(t))^2}_{i}.\underbrace{\frac{\partial v_j(t)}{\partial u_j(t)}}_{ii}
\end{equation}

The portion $(i)$ is always positive. For the portion $(ii)$ choose an increasing monotonic activation function that limits the output of each neuron to a predefined interval. Thus, the two conditions are essential for the dynamic behavior of the Hopfield network to be stable: The matrix ${\bf W} $ must be symmetric and the activation function $g(\cdot) $ must be monotonically increasing.

By establishing the above conditions, then given any set of initial conditions the network will converge to a stable equilibrium point. Then, since the Hopfield network is deterministic, for any initial positions that lie within the region of attraction of a point of equilibrium, the network will always converge asymptotically to this point.

\subsection{Modified Hopfield Artificial Neural Network (mH-NN)}\label{sec:modi}
The neural networks used to solve constrained nonlinear optimization problems are developed using penalty parameters. The equilibrium points corresponding to the solutions of the problem are obtained by choosing appropriate penalty parameters that must be large enough to guarantee the convergence of the network \cite {25}, \cite {26}, \cite {27}.

However, the choice of these parameters is an arduous task and it is usually done through empirical techniques, which may require a very excessive computational effort. In addition, the quality of the final solution also depends on the setting of these parameters.

A detailed analysis of the numerical results in \cite {20} shows that often infeasible results are pointed out as solutions to the problem. In order to overcome problems related to convergence, the authors \cite {24} use a new methodology for the mapping of nonlinear optimization problems called the modified Hopfield network. The modified Hopfield network was implemented in order that the equilibrium points correspond to the solution of the constrained nonlinear optimization problem. The main characteristics of this network are: (i) no weighting constants; (ii) all structural constraints involved in the constrained nonlinear optimization problem are grouped into a single constraint term; (iii) there is no interference between the optimization term and the restriction term and (iv) no initialization parameter is required for simulation execution.

For these problems, a two-term energy function is used:
\begin{equation}\label{funcao_modificada}
E(t)=E^{\rm opt}(t)+E^{\rm conf}(t)
\end{equation}
where
\begin{eqnarray}\label{objetivo_resticao1}
E^{\rm opt}(t)=-\frac{1}{2}{\bf v}(t)^T.{\bf W}^{\rm opt}{\bf v}(t)-{\bf v}(t)^T{\bm \iota}^{\rm opt},
\end{eqnarray}
\begin{eqnarray}\label{objetivo_resticao11}
E^{\rm conf}(t)=-\frac{1}{2}{\bf v}(t)^T.{\bf W}^{\rm conf}{\bf v}(t)-{\bf v}(t)^T{\bm \iota}^{\rm conf},
\end{eqnarray}
the terms $E^{\rm opt}$ and $E^{\rm conf}$ correspond to the optimized energy function and the function that confines all constraints in a single term, respectively. In (\ref{objetivo_resticao1}), the terms ${\bf W}^{\rm opt}$ and ${\bm \iota}^{\rm opt}$ correspond to the optimized weight matrix and the respective bias vector. Finally, the terms ${\bf W}^{\rm conf}$ and ${\bm \iota}^{\rm conf}$ are the weight matrix associated with $E^{\rm conf}$ and the respective bias vector.

The purpose of the network is to simultaneously minimize the energy $E^{\rm opt}(t)$ associated with the objective function of the minimization problem as well as minimizing the energy function $E^{\rm conf}(t)$ involving all constraints of the problem. A simple mapping technique encodes the constraints as terms in the energy function that are minimized when constraints are satisfied \cite{24}, that is:
\begin{eqnarray}\label{objetivo_resticao2}
E(t)= E^{\rm opt}(t)+c_1E_1^{\rm rest}(t)+c_2E_2^{\rm rest}(t)+\ldots+c_kE_k^{\rm rest}(t)
\end{eqnarray}
where $c_i$ are positive constants that give weight to constraints.

The authors of \cite{20} and \cite{24} used the valid subspace technique in order to group all constraints involved in a given problem. {Thus,} the energy function given in (\ref{objetivo_resticao2}) defined by:
\begin{eqnarray}\label{objetivo_resticao3}
E(t)=E^{\rm opt}(t)+c_0E^{\rm conf}(t)
\end{eqnarray}
where $E^{\rm conf}$ confines all restrictions $E_k^{\rm rest}$ of (\ref{funcao_modificada}) in the subspace-valid.

To ensure that $E^{\rm opt}$ optimized when all constraints contained in $E^{\rm conf}$ are satisfied, it involves assigning a high value to the constant $c_0$. This condition makes the network simulation inefficient, since most of the computational effort is to force constraint confinement.

The architecture of the modified Hopfield network is sketched in Fig.\ref{topologia}, where the projection matrix projects the vector ${\bf v}$ into the valid subspace, defined by:
\begin{equation}
{\bf W}^{\rm conf}={\bf I}-{\bm \nabla} h({\bf v})^T \left[{\bm \nabla} h({\bf v}){\bm \nabla} h({\bf v})^T\right]^{-1}{\bm \nabla} h({\bf v}),
\end{equation}
where the function $h$ is defined such that the constraints of the optimization problem can be represented as $h({\bf v})=0$.

Fig. \ref{topologia} represents a suitable example of a recurrent network where the outputs of a neural layer in step (III) are fed back to their inputs in step (I). Indeed, it represents the variable relationship for convergence of the modified Hopfield network whose operating dynamics is implemented through steps (I)-(III).
\begin{figure}[!htbp]
\begin{center}
\includegraphics[trim={1mm 3mm 1mm 6mm},clip,width=.4\textwidth]{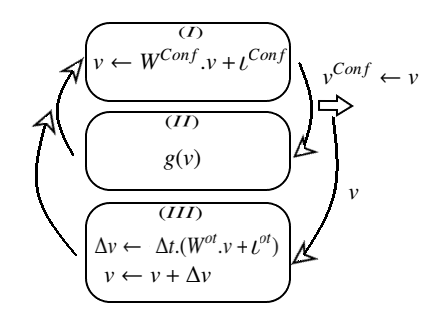}
\caption{Hopfield network for solving constraint optimization problems}\label{topologia}
\end{center}
\end{figure}

The pseudo-code depicted in Algorithm 1 illustrates the basic steps of the mH-NN deployed to solve the OCDMA power allocation problem (\ref{probe}), aiming at finding the minimum transmission power for each user subject to minimum SINR constraint and maximum power budget.

{\begin{algorithm}[!htbp]
\caption{mH-NN for Power Assignment in OCDMA}
\label{Hopfield}
\small
\begin{algorithmic}[1]
  \STATE Initialize: $\p$ with random values\\
  \STATE Introduce: auxiliary variables in the vector $\p$, $\p^*=(\p,q)$, such that $\Gamma(\p)\leq \Gamma^*$ becomes an equality constraint: $h(\p^*)=\Gamma(\p)=\Gamma^*+q=0$.\qquad Also denote $f(\p^*)=J_1(\p)$.  \\
 \STATE Repeat:
 \begin{description}
  \setlength\itemsep{.02mm}
   \item[While] ($\p^+$ do not converge) do
   \begin{description}
     \setlength\itemsep{.05mm}
     \item[Get] the value of $h(\p^+)$
     \item[Get] the Jacobiana matrix ${\bm \nabla} h(\p^+)$
     \item[Update] the value of $\p^+$ from $\p^+ \leftarrow (\p^+-{\bm \nabla} h(\p^+))^T ({\bm \nabla} h(\p^+).{\bm \nabla} h(\p^+)^T)^{-1}{\bm \nabla} h(\p^+)$
     \item[Apply] the activation function
   \end{description}
 \item[\textbf{End of while}]
   \item[Get] the vector ${\bm \iota}^{\rm opt}$ given by ${\bm \iota}^{\rm opt}=-\left[\frac{\partial f(\p^+)}{\partial \p_1^+} \,\, \frac{\partial f(\p^+)}{\partial \p_2^+} \,\, \ldots \,\, \frac{\partial f(\p^+)}{\partial \p_n^+}\right]^T$
   \item[Update] the value of  $\p^+\leftarrow \p^+ +\Delta t({\bf W}^{\rm opt}\p^+ +{\bm \iota}^{\rm opt}).$
   \item[Until] ($\p^+$ stay stationary)
 \end{description}
\STATE End
\end{algorithmic}
\end{algorithm}}

In order to analyze the mH-NN performance, this work evaluate numerically the performance-complexity tradeoff of  three algorithms in solving  the minimum power allocation problem.  In the sequel, we describe the main characteristics of  both nonlinear SQP and classical ALM programming methods.

\subsection{Sequential Quadratic Programming (SQP)}\label{sec:SQP}
The main feature of the sequential quadratic programming method is the determination of the NLP solution as the boundary of the solutions of a quadratic problem sequence.

In our case, the $J(\p)$ function is replaced by a quadratic approximation of the Lagrangian function $\mathcal{L}$, defined below: 
\begin{equation}\label{flagrange}
{\mathcal{L}(\p,\bm{\mu})}=J(\p)+\sum_{i=1}^{K}\mu_i [\Gamma_i^\ast -\Gamma_i(\p)],
\end{equation}
where the nonlinear constraints are replaced by linear approximations thereof. Thus, each iteration of the SQP method solves the following quadratic programming problem:
\begin{eqnarray}\label{eq:SQP}
\max_{\d} & & J(\p^k) +{\bm \nabla} J(\p^k)\tp \d+\frac{1}{2}\d\tp {\bm \nabla}^2{{\p}^k}\mathcal{L}\d \\
{\rm s.t.}& &  -{\bm \nabla} \Gamma_i(\p^k)\tp \d +\Gamma_i^\ast -\Gamma_i(\p^k)\geq0, \quad  i=1\ldots K,\nonumber
\end{eqnarray}
where $\displaystyle {\bm \nabla}^2 {{\p}^k} \mathcal{L} = {\bm \nabla}^2 \mathcal{L}(\p^k) = {\bm \nabla}^2J(\p^k)-\sum_{i = 1}^{K} \mu_i{\bm \nabla}^2 \Gamma_i (\p^k)$.

A more detailed analysis of the construction of the SQP method can be found at \cite{30}. Properties regarding convergence can be found in \cite{39}. The Algorithm \ref{alg:SQP2} describes a pseudo-code for the  SQP method.

\begin{algorithm}[!htbp]
\caption{SQP -- Sequential Quadratic Programming}\label{alg:SQP2}
\small
\begin{algorithmic}[1]
\STATE Choose a starting point $(\p^0,\bm \mu_0)$;
\STATE do $k \leftarrow 0$;
\STATE \textbf{Repeat} \\ 
 Evaluate $J(\p^k), {\bm \nabla} J(\p^k), {\bm \nabla}^2 {\p^k}^2 \mathcal{L}, {\Gamma}_i(\p^k)$ e ${\bm \nabla}{\Gamma_i}(\p^k)$;\\
  Solve \eqref{eq:SQP} to get $\d^k$ and ${\bm\mu}_{k+1}$;
\STATE $\p^{k+1}\leftarrow \mathcal{P}(\p^k+\a_k d^k)$;
where $\mathcal{P}$ is the orthogonal projection operator in the box $p_{\min}\leq \p \leq p_{\max}$.
\STATE \textbf{End(repeat)}
\end{algorithmic}
\end{algorithm}

\subsection{Augmented Lagrangian Method (ALM)}\label{sec:ALM}
The augmented Lagrangian method seeks to solve an NLP in an iterative way, where at each step an optimization problem with simple constraints is solved (in our problem, these constraints define the set $\Omega=\{\p\in\R^{K}p_{\min}\leq p \leq p_{\max}\}$), namely the augmented Lagrangian function defined below for the problem is minimized (\ref{CIR_SNIR}):

%
\begin{equation}
\mathcal{A}_\rho(\p,\bm \mu) =
J(\p)\, +\, \frac{\rho}{2}\, \sum_{i=1}^{K}\left[ [\Gamma_i^\ast - \Gamma_i(\p)]^+ +\frac{\mu_i}{\rho}\right] ^2,
\end{equation}
where $\rho>0$ is the penalty parameter and $\mu_i\geq 0$ are approximations for the Lagrange multipliers.

Generally, in ALM the penalty parameter is set small when starting the method. So hoped that the first iterations favor the achievement of viability.

A more detailed analysis of ALM can be found in \cite{30} and additional results regarding convergence. can be found in \cite{26}, \cite{27}, while extensive results and analysis for this optimization method can be found in \cite{25}. O algorítimo \ref{alg:ALM2} apresenta um pseudocódigo para o ALM.

{\begin{algorithm}[!htbp]
\caption{ALM -- Augmented Lagrangian method}
\label{alg:ALM2}
\small
\begin{algorithmic}[1]
\STATE  {Input Parameters}: 
$\mu_{\max}>0$, \ 
\  $\overline{\mu}_i^1\in[0,\mu_{\max}]\mbox{, }\quad \forall i=1,...,m,$ 
    $\{\varepsilon_k\}\subset \R_+$ 
    so that $\displaystyle \lim_{k\rightarrow\infty}\varepsilon_k=0$.
\STATE do $k\leftarrow 1$
    \STATE \textbf{Repeat} \\
    Calculate $\p^k\in \R^n$ satisfying
    \begin{equation}\label{7.3}
    \left\|\mathcal{P}({\p^k}-{\bm \nabla} \mathcal{A}(\p^k,\overline{\mu}^k))-\p^k\right\| \leq \varepsilon_k,\end{equation}
    where $\mathcal{P}$ is the projection operator in the box $p_{min}\leq \p \leq p_{\max}$.
    \STATE { Update the Lagrange multipliers and the penalty parameter.}
    \STATE \textbf{End(repeat)}
\end{algorithmic}
\end{algorithm}}

\section{Algorithm Implementation, Feasibility and Complexity}\label{sec:implem}
In this subsection relevant aspects on the implementation of the two NLP methods, as well as the proposed modified Hopfield ANN in solving the power allocation problem (\ref{probe}) are developed. The implementation aspects of Algorithm \ref{Hopfield}, \ref{alg:SQP2} and \ref{alg:ALM2}, taking into account realistic OCDMA topologies and considering similar system and channel parameter values,  as in \cite{33}, \cite{34}, \cite{30} are discussed in the next section.

\subsection{Algorithm Implementation Aspects}
In this subsection, the implementation aspects of the previously optimization algorithms taking into consideration realistic optical network topologies are explored. The adopted OCDMA network architecture is the same as at work \cite{41}. The network description and implementation are completely distributed and no training is required. In the following we discuss relevant aspects of implementation for the
three optimization methods applied to PC OCDMA problem.

\subsubsection{SQP Implementation}
The SQP method is initialized deploying a random approximation for the vector $\p$ with the element entries in the range $[p_{\min}; p_{\max}]$,  which is usually adopted in the implementation of the Algorithm \ref{alg:SQP2}. The gradient calculus was implemented by finite differences, while the quadratic subproblem was solved using the interior points (IP) method applied to convex quadratic programs \cite{39}.

\subsubsection{{ALM} Implementation}
The ALM algorithm is initialized with a random approximation for $\p^0\in [p_{\min}; p_{\max}]$, multipliers being zeros, and the penalty parameter value $\rho = 10 $. In addition, if the method repeats the solution of the subproblem with viable points, the algorithm will be interrupted and the convergence attained. Besides, to solve step 3 we used the BFGS quasi-Newton method \cite{39}.

\subsubsection{{mH-NN} Implementation}
The mH-NN algorithm is initialized by deploying a random approximation for $\p^0\in [p_{\min}; p_{\max}]$, following the three steps described in the diagram of Figure \ref{topologia}. $\Delta t = 0.1$ was adopted for distributed implementations and no training requirement.

\subsection{Stopping Criterion and Feasibility}
The same stopping criterion has been considered for all power allocation algorithms analyzed. Hence, if after the $k$th external iteration,  $\textbf{p}^k$ results feasible, and the value
\begin{equation}\label{criterio_parada}
 \xi =\,  \parallel\textbf{p}^{k} - \textbf{p}^{k-1} \parallel \,\, <\, 10^{-6},
\end{equation}
the algorithm stops execution, reaching convergence.

The \emph{feasibility} in the context of OCDMA power control problem is considered as: 
\begin{equation}\label{factibilidade}
\mathcal{F}_i^{(k)} = \, [\Gamma_i^* - \Gamma_i]^+ = \max \{0, |\Gamma_i^* - \Gamma_i|\} \qquad  \text{if } \quad p_{\min}\leq \p^k\leq p_{\max}, \qquad \forall k \,\,\,\text{ iteration}
\end{equation}
Thus, the null value of the feasibility $\mathcal{F}^{(k)} =\max\{\mathcal{F}_1^{(k)}, \mathcal{F}_2^{(k)},\ldots,\mathcal{F}_K^{(k)}\}$ at the $k$th iteration indicates that power vector $\p^k$ satisfying the constraints of the problem (\ref{probe}). %

\subsection{Normalized Mean Squared Error (NMSE)}
The quality of the solution reached by the power allocation algorithms in an iteration can be measured by the degree of proximity to the optimal solution $\p^*$, being quantified through the normalized mean square error (NMSE) when the equilibrium is reached. The quality of the solution achieved by an specific algorithm in solving problem (\ref{probe}) is simply  defined by \cite{17}:
\begin{equation}\label{NMSE}
\textsc{nmse} = \mathbb{E}\left[\frac{||\p_t[n]-\p^*||^2}{||\p^*||^2}\right],
\end{equation}
where $||\cdot||^2$ denotes the squared Euclidean distance between vector $\p_t[n]$ to the optimum solution vector $\p^*$ at the $n$-th iteration of the $t$-th realization and $\mathbb{E}[.]$ is the expectation operator. This measure will also be analyzed in the context of this work.

\subsection{Algorithm Robustness}
The algorithm robustness $\mathcal{R}$ can be thought as the ratio between the number of convergence success $\texttt{cS}$  to the total number of process realizations $\mathcal{T}$:
\begin{equation}\label{R}
\mathcal{R} = \frac{\texttt{cS}}{\mathcal{T}}.100 \ \ [\%]
\end{equation}
The convergence success event is confirmed when the stopping criterion and feasibility are achieved.

\section{Numerical Results}\label{sec:resu}
In this section the objective is to analyze through numerical tests the behavior of the modified Hopfield neural network for the resolution of the minimum power allocation problem in \eqref{probe}, aiming at improving the overall energy efficiency of the OCDMA system. Also, we compared it with the nonlinear programming methods, and taking into consideration the realistic optical network topologies based on 2-D {\it multiple-length extended wavelength hopping prime code} (MLEWHPC) \cite{41}.

The  MLEWHPC codes are composed by a set of 2-D multiple-length constant-weight EWHPCs. Such code set is able to support a large variety of multimedia services, such as data, voice, image and video, while accommodating simultaneously all kinds of subscribers with very different bit-rates and quality-of-service (QoS). These codes have ideal correlation properties which can be obtained by extending wavelength-hopping prime codes with single length. The resulting MLEWHPCs present identical autocorrelation peaks and low cross-correlation values of at most one \cite{41}, \cite{42}, \cite{43}.

We apply the three algorithms for different numbers of nodes of an OCDMA network considering the parameter values summarized in Table \ref{tab:param3}. The adopted target SNIR of $20$ dB has been chosen to achieve suitable transmission in a single rate network, resulting in a BER of less than the free limit of error $(\sim10^{-12})$. Besides, the choice of network size considered for numerical analysis in this section takes into account two scenarios: \\
\underline{Scenario A} represents a medium system loading with $K = \{8, 16, 32 \}$ users;\\
\underline{Scenario B} is more challenging high loading optical network in considering $K =\{48, 64, 128\}$ users and different levels of QoS.

\begin{table}[!htbp]
\caption{Adopted Parameter Values}
\vspace{-4mm}
\small
\centering
{{\renewcommand{\arraystretch}{.75}%
\resizebox{0.6\textwidth}{!}{
\begin{tabular}{llr}
\hline
\bf Parameter & \bf Value & \bf Unit\\
\hline
\hline
Modulation order & $M=2$ & (Binary)\\
Transponder Inefficiency & $\iota = 2.7$ & [W/Gbps]\\
White Noise std & $ \sigma = 0.032$&  [dB]\\
Planck constant &$h=6.63 \times 10^{-34}$ & [J/Hz]\\
Chip Period & $T_c=9$  &[ps]\\
Link length & $[4;100]$ &[km]\\
Max. Laser power & $p_{\max} = 20$ & [dBm] \\
Min. Laser power & $p_{\min} = (p_{\max}-90)$ & [dBm] \\
\hline
\multicolumn{3}{c}{\bf Scenario A: \quad single rate}\\
\hline
Number of users & $K\in\{8, 16, 32\}$ & [users] \\
Min. user rate & $r_{\min}^{\rm serv}= 30$ & [Mbps] \\
Sequence length &  $F_i=\frac{T_b}{T_c} =  121$ & \\
Max. BER acceptable & $\textsc{ber}_{\max} \leq 10^{-12}$ & \\
SNIR target (min) & $\Gamma^*=20$ & [dB]\\
\hline
\multicolumn{3}{c}{\bf Scenario B: \quad  single rate,\, different QoS}\\
\hline
Number of users & $K\in\{48, 64, 128\}$ & [users]\\
Class I:  \,\,\,\,\quad SNIR$^{\rm (I)}$ and  & $\Gamma^{{\rm (I)}*} = 17$ & [dB]\\
Class II:  \,\,\,\,\,  SNIR$^{\rm (II)}$ & $\Gamma^{{\rm (II)}*} = 20$ &[dB]\\
Class III:  \quad SNIR$^{\rm (III)}$ & $\Gamma^{{\rm (III)}*} = 22$ & [dB]\\
Min. user rate per class & $r_{\min}^{\rm serv}=25, \, 30$ or $35$ & {[Mbps]}  \\
\hline\hline
\multicolumn{3}{c}{\textbf{\textit{Algorithm Initialization and Convergence}}}\\
\hline
Power vector initialization &   \multicolumn{2}{l}{$\p^0\sim\mathcal{U}[p_{\min}; p_{\max}]$}\\
Max. number of iterations & \multicolumn{2}{l}{$\mathcal{I}=10$ or $15$ (under pertubation)}\\
Convergence criterion &  \multicolumn{2}{l}{$\mathcal{F} \leq 10^{-4}$ and $ \xi  <10^{-6}$, \, $\textbf{p}$ feasible}\\
\hline
\end{tabular}
}}}
\label{tab:param3}
\end{table}

The simulation results were performed using the software MatLab $8.0$ running under Windows $10$ Home Language,  version $1803$, Intel(R)Core processor  i5-8250U @CPU 1.60GHz, 8.00GB of RAM and 64-bit operating system.
\subsection{Power Assignment Optimization (Scenario A)}
For the first scenario, the three algorithms addressed took into account $K=8, 16$ or $32$ users of an OCDMA network whose parameters are described in Table \ref{tab:param3}. Besides, Table \ref{tabelares} reveals the numerical values referring to the performance of the three power allocation methods, including execution time, minimum value of sum-power $J_1(\textbf{p})$, number of iterations required for convergence, measure of feasibility $\mathcal{F}$, number of {\it floating-point operations per second} (\textsc{flops}), normalized MSE, eq. \eqref{NMSE}, and sum-rate $\sum_{i=1}^K r_i$, with $r_i \geq r_{\min}^{\rm serv}$. The flops were obtained through an adaptation of Hang Qian's Contour FLOPS program\footnote{Available for download at {\tiny \url{www.mathworks.com/matlabcentral/fileexchange/50608-counting-the-floating-point-operations-flops}}}.

Table \ref{tabelares} shows that all methods have converged to feasible points and also that the three methods reach the optimum power allocation values for Scenario A. The best feasibility levels were obtained by the SQP and modified Hopfield NN methods and the best NMSE values were obtained by the mH-NN method, with advantage for mH-NN method when $K$ increases. For $8$ and $16$ users, Hopfield and SQP had close amounts of FLOPS, while Hopfield was slightly faster than SQP considering the execution time, specifically for higher problem dimension, i.e., for $K\geq 32$, SQP users reached lower number of FLOPS. However, mH-NN method has achieved convergence by requiring shorter processing time. It can be observed that Hopfield and SQP performed very closely while ALM consumed a greater number of FLOPS and thus consumed more time for convergence. Finally, the SQP and Hopfield methods attain practically the same levels of sum-rate, i.e., $\sum_{i=1}^Kr_i \geq K\cdot r_{\min}^{\rm serv}$, for all user values while the ALM method presented little difference to the values of sum-rate due to its marginal feasibility performance degradation when compared to the other two methods. Despite the adopted constrain $r_{\min}^{\rm serv} = 30$ Mbps, the laser power budget is enough to attain an average per user rate of $\bar{r}_{i} = 34.34$ Mbps.

\begin{center}
\begin {table}[htbp!]
\small
\centering
\caption {Execution time, the number of external iterations for convergence, feasibility for the three algorithms, FLOPS, NMSE and sum-rate considering the increase in the number of optical nodes $K$. The average per user rate has resulted $\bar{r}_{i} = 34.34$ Mbps.}
\label{tabelares}
{{\renewcommand{\arraystretch}{1.2}%
\resizebox{0.80\textwidth}{!}{%
\begin {tabular}{|c|c|c|c|c|c|c|c|c|c|}
\hline \bf Method & {\bf Time} [sec.] & $J_1(\textbf{p})$ [W] & \bf Iterations  & \bf Feasibility  $\mathcal{F}$& \bf \small \footnotesize FLOPS& {\bf \footnotesize NMSE} & $\sum_{i=1}^K r_i$ [Mbps]\\ \hline\hline
\multicolumn{7}{|c|}{\bf 8 users}\\
\hline ALM        &   0.34078        &  7.90100e-05     &   {6}    &   {3.45931e-05}  &   2.9615e+5   &  0.49289      &  2.74721e+2   \\
\hline SQP        &   0.01577        &  7.90197e-05     &   {3}    &   \underline{2.74802e-14}  &   6.5850e+4   &  0.25633      &  2.74751e+2  \\
\hline mH-NN      &   0.01215        &  7.90105e-05     &   {2}    &   {4.78985e-06}  &   2.8885e+4   &  2.28502e-12  &  2.7475e+2  \\
\hline
\multicolumn{7}{|c|}{\bf 16 users}\\
\hline ALM        &   0.59639       &  4.70084e-04      &   {6}     &  {1.82361e-05}   &  1.4106e+6    &  0.37642      & 5.5970e+2    \\
\hline SQP        &   0.01965       &  4.70114e-04      &   {3}     &  \underline{3.40500e-11}   &  4.1945e+5    &  0.33332      & 5.4950e+2   \\
\hline mH-NN      &   0.01559       &  4.70094e-04      &   {2}     &  {8.66522e-06}   &  2.0925e+5    &  1.49264e-11  & 5.4942e+2    \\
\hline
\multicolumn{7}{|c|}{\bf 32 users}\\
\hline ALM        &   0.98670       &  0.01869          &   {6}     &  {2.7581e-06}    &  2.6283e+7    &  0.30801      & 1.0989e+3 \\
\hline SQP        &   0.34545       &  0.01869          &   {3}     &  {1.1180e-10}    &  2.9736e+6    &  0.22872      & 1.0990e+3  \\
\hline mH-NN      &   0.01508       &  0.01869          &   {2}     &  \underline{9.8134e-18}    &  3.2326e+6    &  4.44322e-07  & 1.0990e+3 \\
\hline
\end {tabular}
}}}
\end {table}
\end{center}

Fig. \ref{evolucao1} depicts the convergence evolution of the individual power levels for the three power allocation methods in solving (\ref{probe}) in comparison to the inversion matrix (Tarhuni) solution obtained by (\ref{tarhuni}). In Figure \ref{evolucao1}(a) one can observe that the ALM method starts to approach the individual power levels of the solution after the fourth iteration and the convergence occurs in the sixth iteration. While the numerical convergence results for the SQP method in Fig.\ref{evolucao1}(b) reveals that the optimal power levels are simultaneously attained early in the second iteration, satisfying the convergence criteria in the third iteration. Finally, Fig.\ref{evolucao1}(c) shows the mH-NN method reaching the solution right after the first iteration and convergence criteria in the second iteration.

\begin{figure}[!htbp]
\begin{center}
\small
\includegraphics[width=0.92\textwidth]{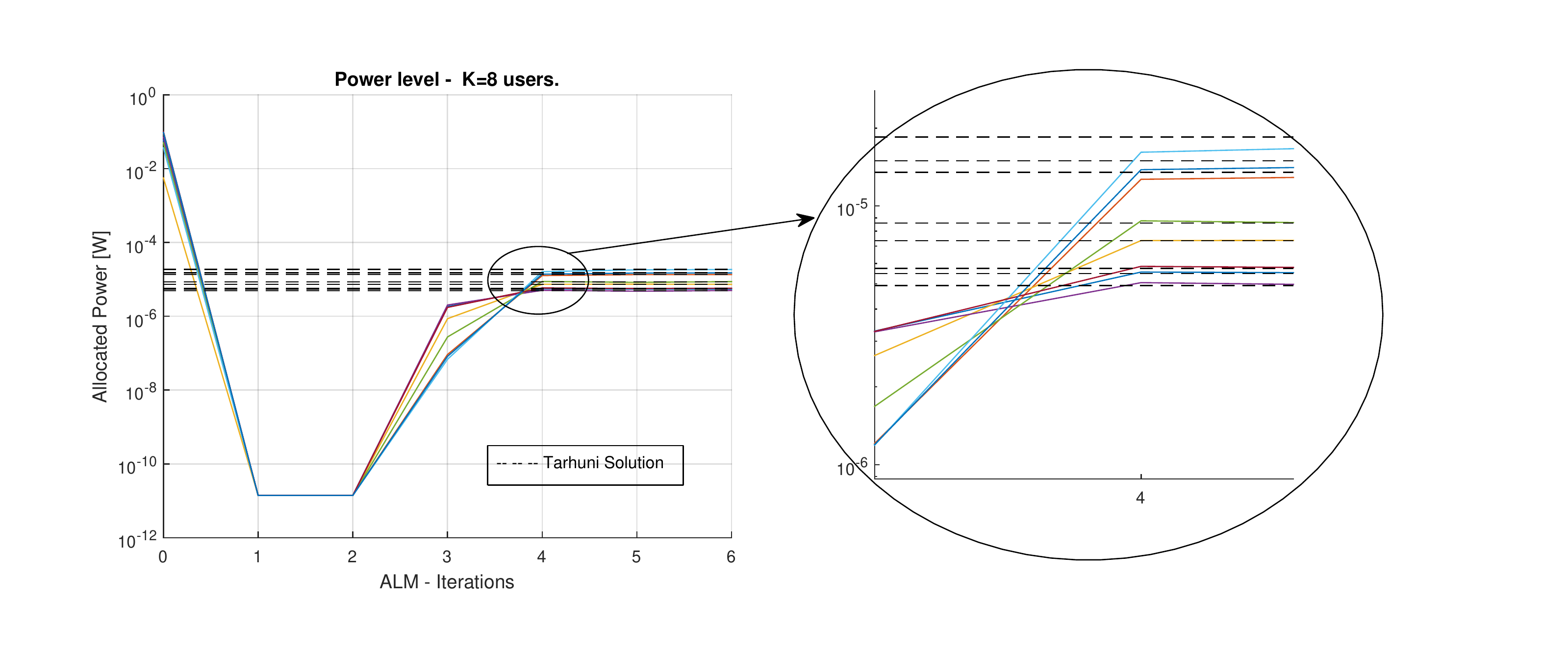}\\
a) ALM algorithm
\includegraphics[width=0.98\textwidth]{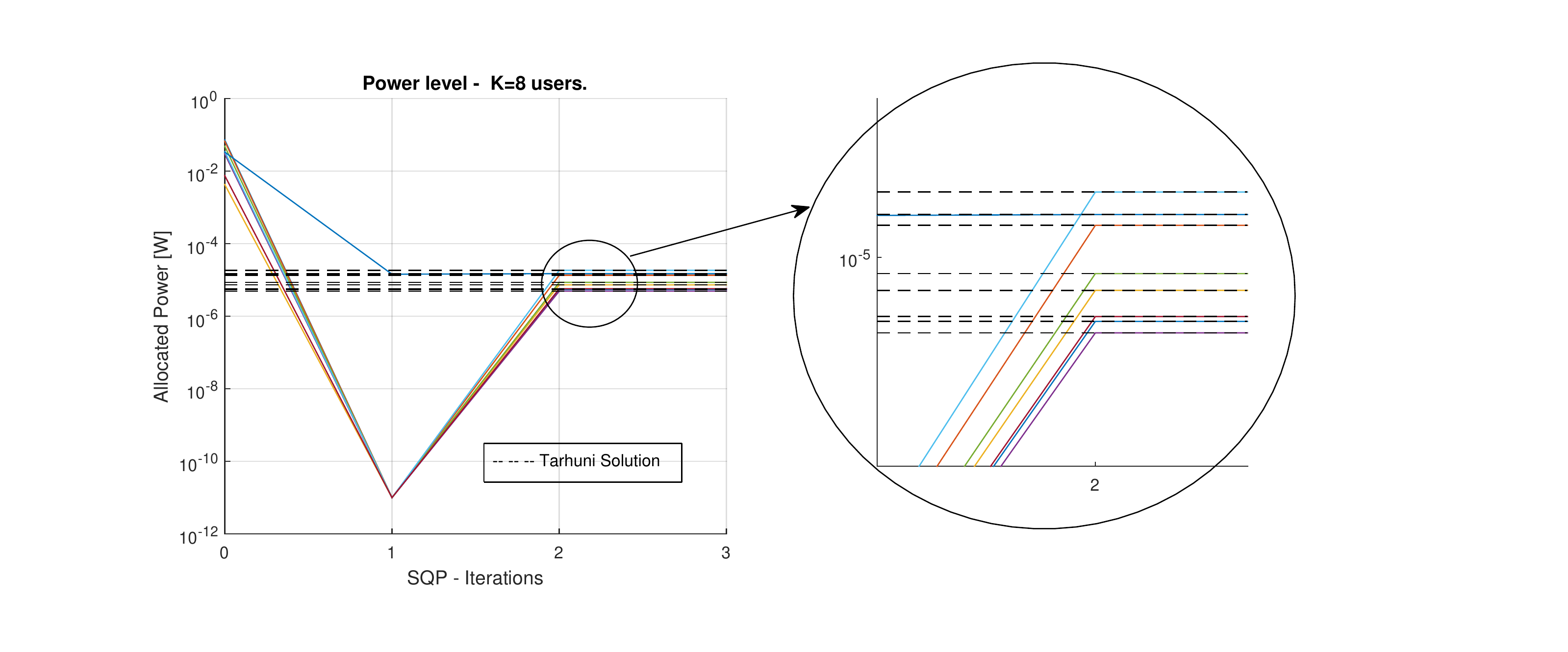}\\
a) SQP algorithm
\includegraphics[width=0.92\textwidth]{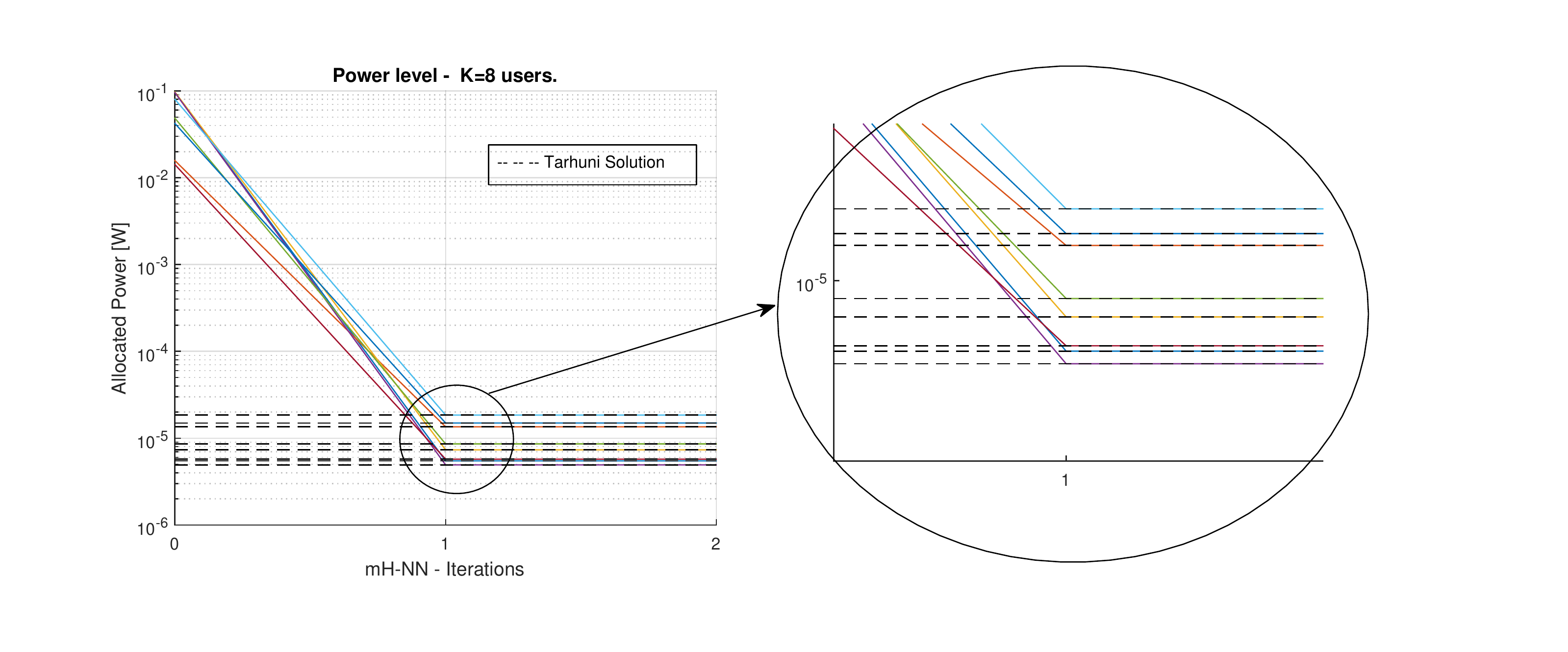}\\
a) mH-NN algorithm
\caption{Power evolution until the equilibrium in Scenario A, $K=8$ users: a) ALM; b) SQP; c) mH-NN algorithms}\label{evolucao1}
\end{center}
\end{figure}

Fig. \ref{SumK-R_8,16,32}.(a) presents the evolution of sum-power $J_1({\bf p})$ along the iterations. It may be noted that regardless of the number of users, mH-NN algorithm reaches the required minimum power assignment as early as the first iteration. On the other hand, SQP approaches the convergence after the second iteration and ALM begins to approach the optimal power solution after the third iteration. Complementary, Fig. \ref{SumK-R_8,16,32}.(b) depicts the evolution of the sum of the user rates along with the iterations for the three methods discussed. As a result, the behavior is similar to the sum-power of the corresponding graphs (for the same $K$ network loading) in  Fig. \ref{SumK-R_8,16,32}.(a). One can see that the ALM and SQP methods, at the beginning of iterations, get very different values from the optimal value of sum-rate.

\begin{figure}[!htbp]
\centering
\includegraphics[trim={1mm 1mm 1mm 1mm},clip,width=.97\textwidth]{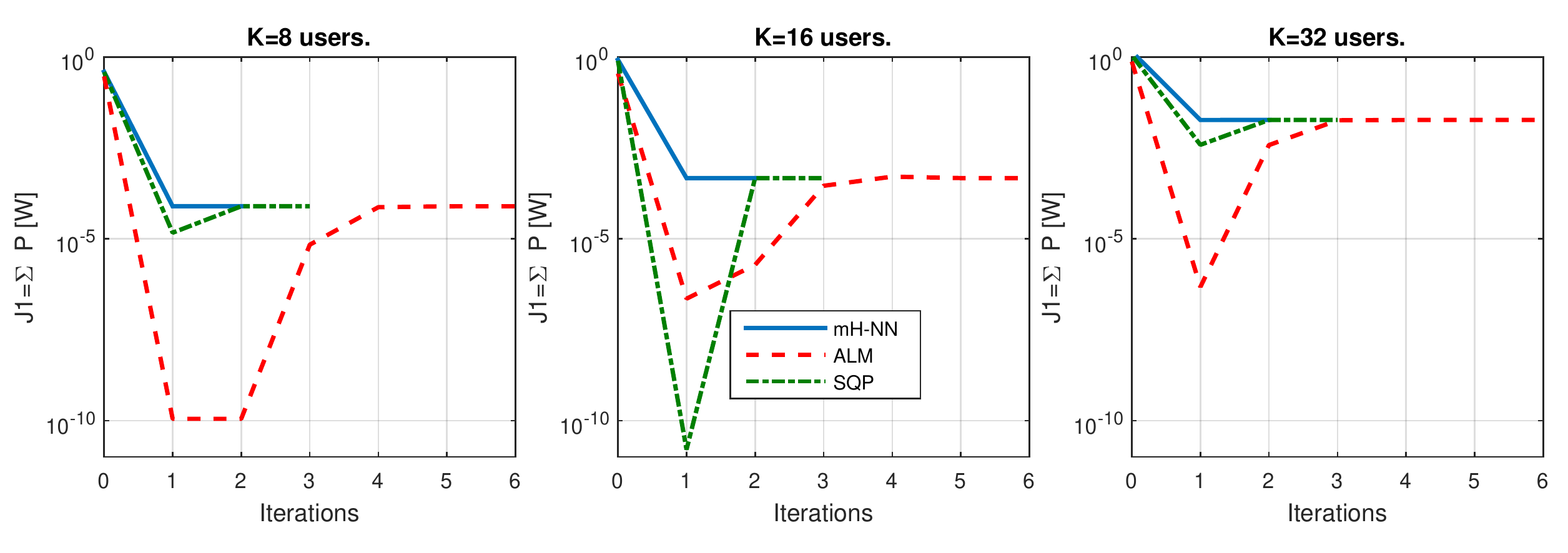}
(a) Sum-Power levels
 \includegraphics[trim={1mm 1mm 1mm 1mm},clip,width=.97\textwidth]{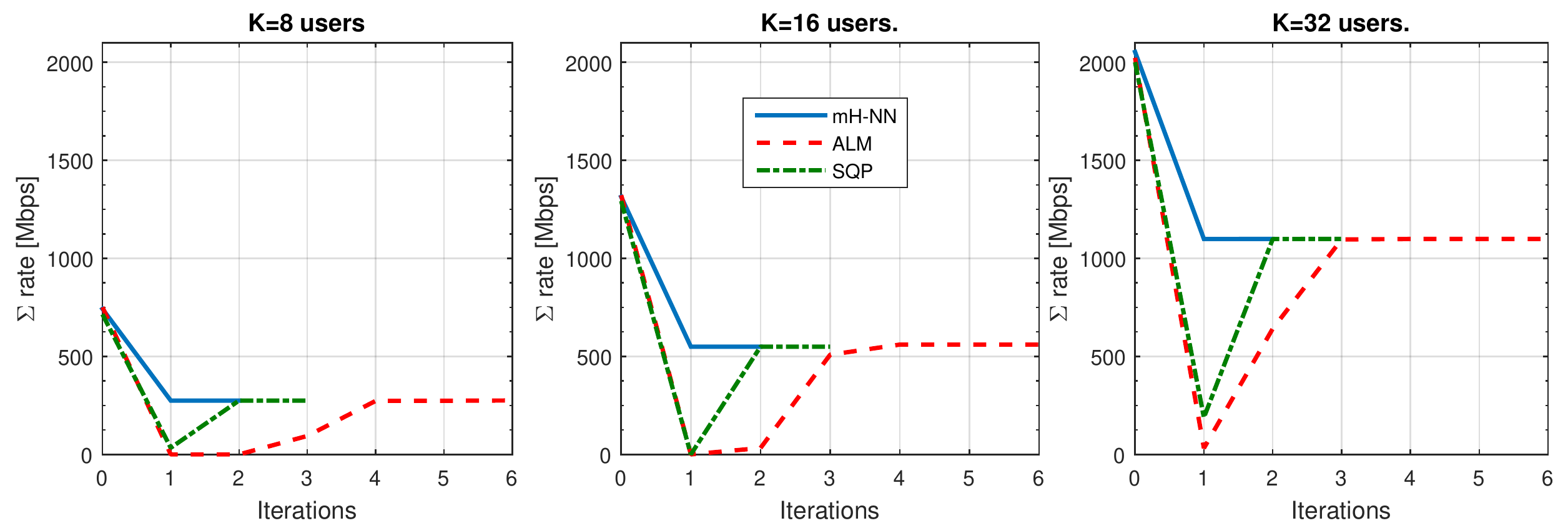}
 (b) Sum-Rate levels
\caption{Sum-Power (a) and Sum-Rate (b) allocation across the iterations for  $K=8, \ 16$ and $32$ users in Scenario A.}
\label{SumK-R_8,16,32}
\end{figure}

A deeper analysis of the numerical convergence results in Fig.\ref{evolucao1} and Fig. \ref{SumK-R_8,16,32}.(a) evidences that nonlinear programming methods in the first iterations may preferentially seek better values of the objective function over feasibility. This behavior is described in the literature as the voracity in reducing the objective function magnitude, which has already been reported in \cite{35} for the ALM method, but considering the dimension of the networks treated in the Scenario A and such characteristic did not affect the convergence of the method herein. However, we can highlight that ALM demanded a greater number of iterations to achieve convergence when compared to the SQP method. On the other hand, the modified Hopfield neural network (mH-NN) method has demonstrated a distinct behavior regarding NLP methods, approaching convergence very soon, typically after the first iteration.

\newpage
\subsection{Power Assignment Optimization with Different Levels of QoS (Scenario B)}
In this subsection, the minimum power allocation problem is defined for a single-rate different QoS OCDMA system, considering the parameters previously described in Table \ref{tab:param3} for Scenario B and larger networks with $K =48$, $64$ and $128$ users. It also has analyzed three different levels of QoS determining distinct classes of QoS, which are associated to different attainable single-rate OCDMA systems, namely Class I, II and III, and defined by the following SNIR:
$$\Gamma^*=17 {\rm dB}\quad (r_{\min}^{\rm(\textsc{i})}); \qquad  20 {\rm dB}\quad  (r_{\min}^{\rm(\textsc{ii})}); \qquad {\rm and} \qquad 22 {\rm dB} \quad (r_{\min}^{\rm(\textsc{iii})})
$$
As in the Scenario A, herein the three power assignment algorithms are compared in terms of execution time, minimum power solution, number of iterations, feasibility, FLOPS, NMSE and attainable data rates, as  depicted in Table \ref{tabelaresB1-B3}.
It can be noted that in Scenario B, the three methods maintained similar performance for the three classes of service (single-rate), Class I, II and III to that obtained in Scenario A, where SQP and mH-NN methods present similar execution time values while attain the best levels of feasibility. Moreover, the mH-NN is able to attain better NMSE levels due to the fact that already in the first iteration the method has been able to achieve suitable approximations for the solution even under high loading systems of $K=128$ users. In turn, the ALM presented the worst values for NMSE because it consumes greater number of iterations to achieve the same solution quality. Considering the attainable sum-rate, one can observe the behavior similar to Scenario A, where SQP and the modified Hopfield (mH-NN) present similar values and ALM a minimum difference, probably due to its inferior performance regarding the feasibility.
Notice that in solving the minimum optimal power allocation ($\p^*$) problem with different level of QoS, one can observe that the value of $\Gamma_i$ coincides with $\Gamma_i^*$, so the value of CIR for each user is minimal. On the other hand, after convergence (using $\p^*$) one can observe that sum-rate is slightly greater than $ K \cdot r_{i,\min}$. This result reveals that the minimum power vector found takes into consideration all the system impairments while the calculation of the average rate per user does not take such factors into account.

\begin {table}[!htbp]
\small
\centering
\caption {Class I, II and III --  Execution time, sum-power, number of iterations for convergence, feasibility, FLOPS, NMSE and sum-rate for  $K=48, 64, 128$  optical nodes.}
\label{tabelaresB1-B3}
{\footnotesize  {\bf (a) Class I} \quad --\quad  $\Gamma^*=17 {\rm dB}\quad (r_{\min}^{\rm(\textsc{i})} =25 {\rm Mbps})$ and $\bar{r}_{i}^{\rm(\textsc{i})} \approx 29.49$ Mbps.}
{{\renewcommand{\arraystretch}{1.2}%
\resizebox{0.75\textwidth}{!}{%
\begin {tabular}{|c||c|c|c|c|c|c|c|c|}
\hline \bf Method & {\bf Time} [sec.] & $J_1(\p)$ [W] & \bf Iterations & \bf Feasibility $\mathcal{F}$ &{\bf {FLOPS}} &\bf NMSE &$ \sum_{i=1}^K r_i$ [Mbps] \\ \hline\hline
\multicolumn{8}{|c|}{\bf 48 users}\\
\hline ALM        &   1.21379        &  0.026371                     &   {6}         &   {1.24862e-06}    &   3.5609e+8  &  0.36453    & 1.4155e+3  \\
\hline SQP        &   0.03135        &  0.026371                     &   {4}         &   {7.06701e-13}    &   9.5138e+6  &  0.14921    & 1.4155e+3   \\
\hline mH-NN   &   0.03266        &  0.026371                     &   {3}         &   {5.48562e-18}    &   2.2936e+7  &  2.67154e-09& 1.4155e+3  \\
\hline
\multicolumn{8}{|c|}{\bf 64 users}\\
\hline ALM         &   1.65431       &  0.035138                     &   {6}         &  {1.38161e-06}     &   9.4509e+8  &  0.35719    & 1.8874e+3   \\
\hline SQP         &   0.08348       &  0.035138                     &   {4}         &  {1.61612e-12}     &   2.2146e+7  &  0.24360    & 1.8874e+3  \\
\hline mH-NN    &   0.05468       &  0.035138                     &   {3}         &  {6.49071e-18}     &   5.7321e+7  &  5.33412e-09& 1.8874e+3 \\
\hline
\multicolumn{8}{|c|}{\bf 128 users}\\
\hline ALM         &   2.02370       &  0.13220                      &   {7}         &  {9.88553e-05}      &   3.9875e+9 &  0.39053    & 3.7748e+3\\
\hline SQP         &   0.15053       &  0.13200                      &   {3}         &  {1.53822e-12}      &   1.7241e+8 &  0.05997    & 3.7748e+3    \\
\hline mH-NN    &  0.34171       &  0.13200                      &   {3}          &  {6.490744e-18}     &   4.8104e+8 &  3.19617e-09& 3.7748e+3   \\
\hline
\end {tabular}
}}}
\vspace{4mm}

{\footnotesize  {\bf (b) Class II} \quad --\quad  $\Gamma^*=20 {\rm dB}\quad (r_{\min}^{\rm(\textsc{ii})} =30 {\rm Mbps})$ and $\bar{r}_{i}^{\rm(\textsc{ii})} \approx 34.34$ Mbps.}
{{\renewcommand{\arraystretch}{1.2}%
\resizebox{0.75\textwidth}{!}{%
\begin {tabular}{|c|c|c|c|c|c|c|c|c|}
\hline \bf Method & {\bf Time} [sec.] & $J_1(\textbf{p})$ [W] & \bf Iterations & \bf Feasibility $\mathcal{F}$  & {\bf {FLOPS}}  &  NMSE  & $\sum_{i=1}^K r_i$ [Mbps] \\ \hline\hline
\multicolumn{8}{|c|}{\bf 48 users}\\
\hline ALM        &   1.73390        &  0.05262                  &   {6}           &{1.27104e-06}    & 4.3685e+8   &  0.33348      &   1.6485e+3 \\
\hline SQP        &   0.03134        &  0.05262                  &   {3}           &{7.39512e-12}    & 9.5230e+6   &  0.10153      &   1.6485e+3      \\
\hline mH-NN   &   0.03260        &  0.05262                  &   {3}           &{1.15068e-17}    & 2.4285e+7   &  6.49577e-09  &   1.6485e+3   \\
\hline
\multicolumn{8}{|c|}{\bf 64 users}\\
\hline ALM         &  1.91301       &  0.07011                   &   {6}           &{6.95347e-04}    & 1.1431e+9  & 0.33092        &   2.1980e+3   \\
\hline SQP         &  0.08343       &  0.07012                   &   {3}           &{1.50302e-11}    & 2.2149e+7  & 0.14114        &   2.1980e+3  \\
\hline mH-NN    &  0.05468       &  0.07012                   &   {3}           &{1.04083e-17}    & 5.7321e+7  & 1.51221e-08    &   2.1980e+3     \\
\hline
\multicolumn{8}{|c|}{\bf 128 users}\\
\hline ALM         &   2.48559       &  0.26679                  &   {7}           &{5.58532e-04}    &  4.9978e+9  &  0.39589      &   4.4017e+3 \\
\hline SQP         &   0.15053       &  0.26345                  &   {3}           &{7.52932e-12}    &  1.7534e+8  &  0.03146      &   4.3960e+3  \\
\hline mH-NN    &   0.17014       &  0.26345                  &   {3}           &{1.09713e-17}    &  4.5572e+8  &  3.73725e-09  &   4.3960e+3  \\
\hline
\end {tabular}
}}}

\vspace{4mm}

{\footnotesize  {\bf (c) Class III} \quad --\quad  $\Gamma^*=22 {\rm dB}\quad (r_{\min}^{\rm(\textsc{iii})} =35 {\rm Mbps})$ and $\bar{r}_{i}^{\rm(\textsc{iii})} \approx 37.60$ Mbps.}
{{\renewcommand{\arraystretch}{1.2}%
\resizebox{0.75\textwidth}{!}{%
\begin {tabular}{|c|c|c|c|c|c|c|c|c|}
\hline \bf Method & {\bf Time} [sec.] & $J_1(\textbf{p})$ [W] & \bf Iterations  & \bf Feasibility $\mathcal{F}$   &  {\bf Flops} &\bf NMSE & $\sum_{i=1}^K r_i$ [Mbps] \\ \hline\hline
\multicolumn{8}{|c|}{\bf 48 users}\\
\hline ALM        &   1.05472        &  0.08342                     &   {6}     &   {3.55791e-07}  & 3.4363e+8  &  0.31702     &   1.8046e+3 \\
\hline SQP        &   0.03901        &  0.08342                     &   {3}     &   {2.41023e-11}  & 9.5851e+6  &  0.05154     &   1.8046e+3     \\
\hline mH-NN   &   0.03052        &  0.08342                     &   {3}     &   {1.20122e-17}  & 2.2936e+7  &  7.56132e-09 &   1.8046e+3     \\
\hline
\multicolumn{8}{|c|}{\bf 64 users}\\
\hline ALM         &   1.420707       &  0.11116                     &   {6}    &  {1.43765e-06}  & 9.7314e+8  &  0.30859      &   2.4061e+3  \\
\hline SQP         &   0.098363       &  0.11116                     &   {3}    &  {7.34822e-11}  & 2.1428e+7  &  0.08301      &   2.4061e+3 \\
\hline mH-NN    &   0.105133       &  0.11116                     &   {3}    &  {1.551583e-17} & 5.4137e+7  &  2.56962e-08  &   2.4061e+3       \\
\hline
\multicolumn{8}{|c|}{\bf 128 users}\\
\hline ALM         &   2.734116      &  0.41761                      &   {7}    &  {2.36749e-04}   & 4.8290e+9  &  0.31848     &   4.8122e+3  \\
\hline SQP         &   0.215497       &  0.41761                     &   {3}    &  {4.23523e-11}   & 1.7381e+8  &  0.01831     &   4.8122e+3 \\
\hline mH-NN    &   0.242005       &  0.41761                     &   {3}    &  {1.55134e-17}   & 4.5573e+8  & 5.34532e-09  &   4.8122e+3 \\
\hline
\end {tabular}
}}}
\end {table}


Figure \ref{K48,64,128_17_20_22db} shows the evolution of sum-power along the external iterations for the three power allocation algorithms operating under Class I, II and III OCDMA system, respectively. Indeed, considering Fig. \ref{K48,64,128_17_20_22db}(a), it can be noted that regardless of the number of users in the high loading Scenario B ($K\geq48$ users), the SQP and mH-NN algorithms are able to reach values very close to the minimum required power in the second iteration, while the ALM method again presents difficulties in the first iterations to attain convergence due to its voracity feature, as already seen in Scenario A. Besides, inspecting Fig. \ref{K48,64,128_17_20_22db}(b) and \ref{K48,64,128_17_20_22db}(c), depicting the evolution of sum-power and sum-rate
over iterations to Class II and III, respectively, one can note that regardless of the number of users that SQP and mH-NN perform very closely in terms of number of iterations towards the convergence. The same trend is confirmed in terms of sum-rate for this scenario; Fig. \ref{K48,64,128_17_20_22sumrate} exhibits the evolution of sum-rate with respect to the iterations for Class I, II and III.

\begin{figure}[!htbp]
\centering
\small
\includegraphics[trim={1mm 1mm 1mm 1mm},clip,width=1\textwidth]{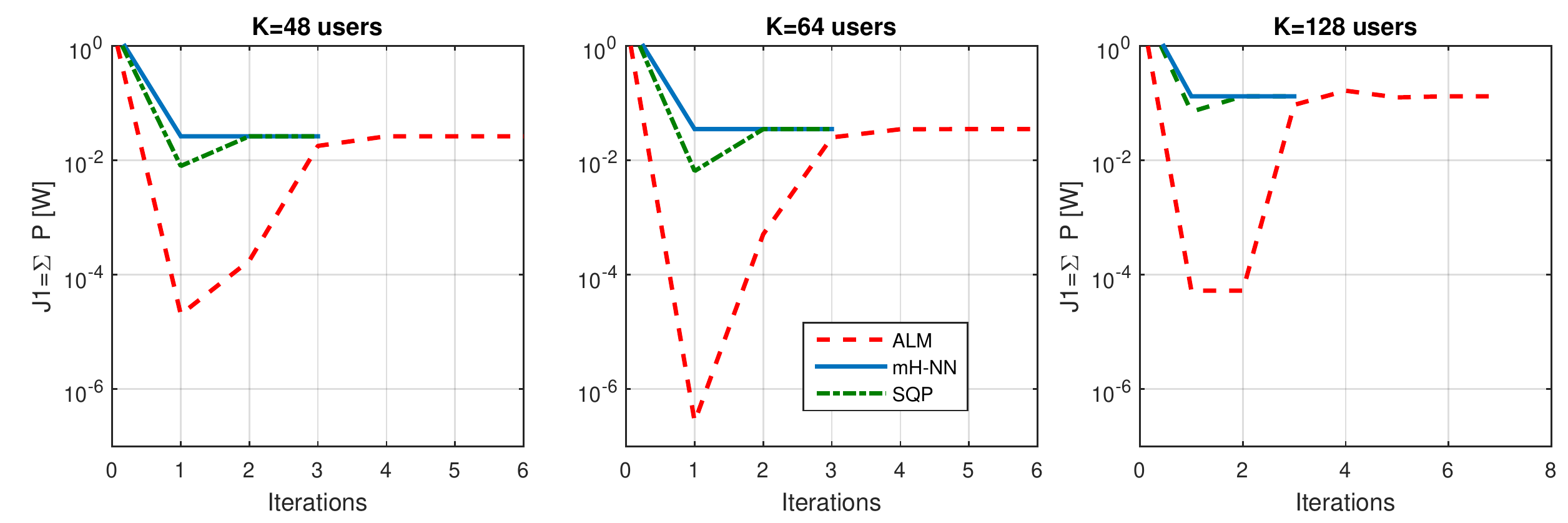}
(a) $\Gamma^*= 17$dB\\
\vspace{4mm}
\includegraphics[trim={0mm 1mm 1mm 1mm},clip,width=1\textwidth]{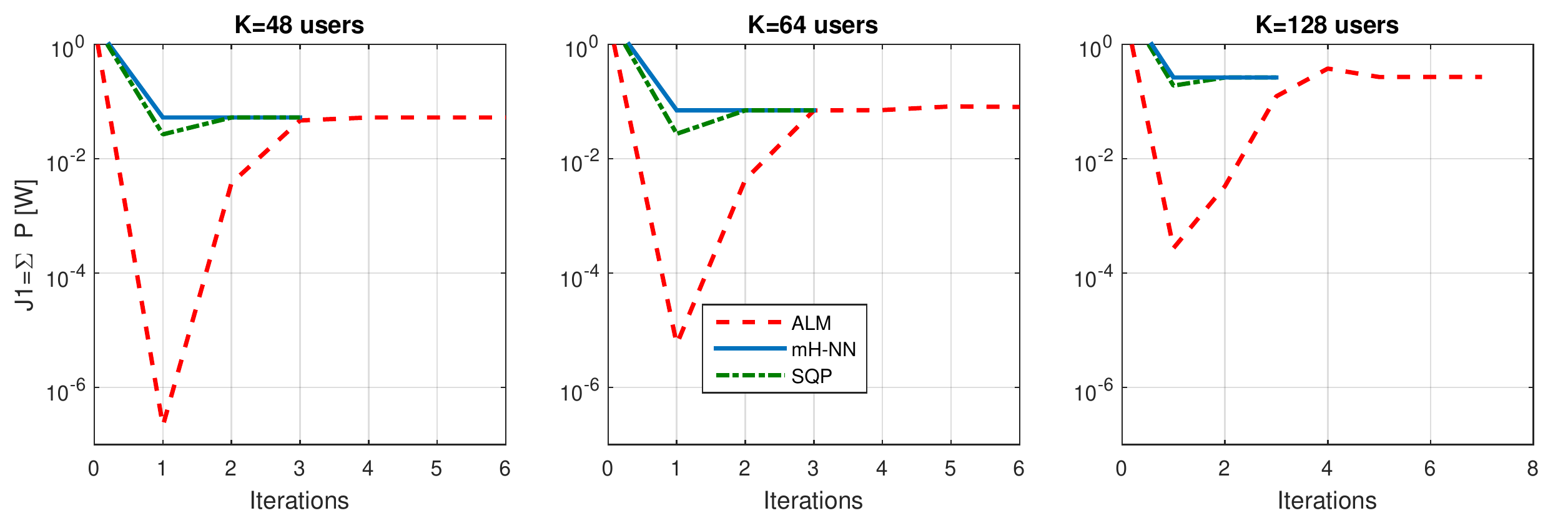}
(b) $\Gamma^*= 20$dB\\
\vspace{4mm}
\includegraphics[trim={2mm 1mm 1mm 1mm},clip,width=1\textwidth]{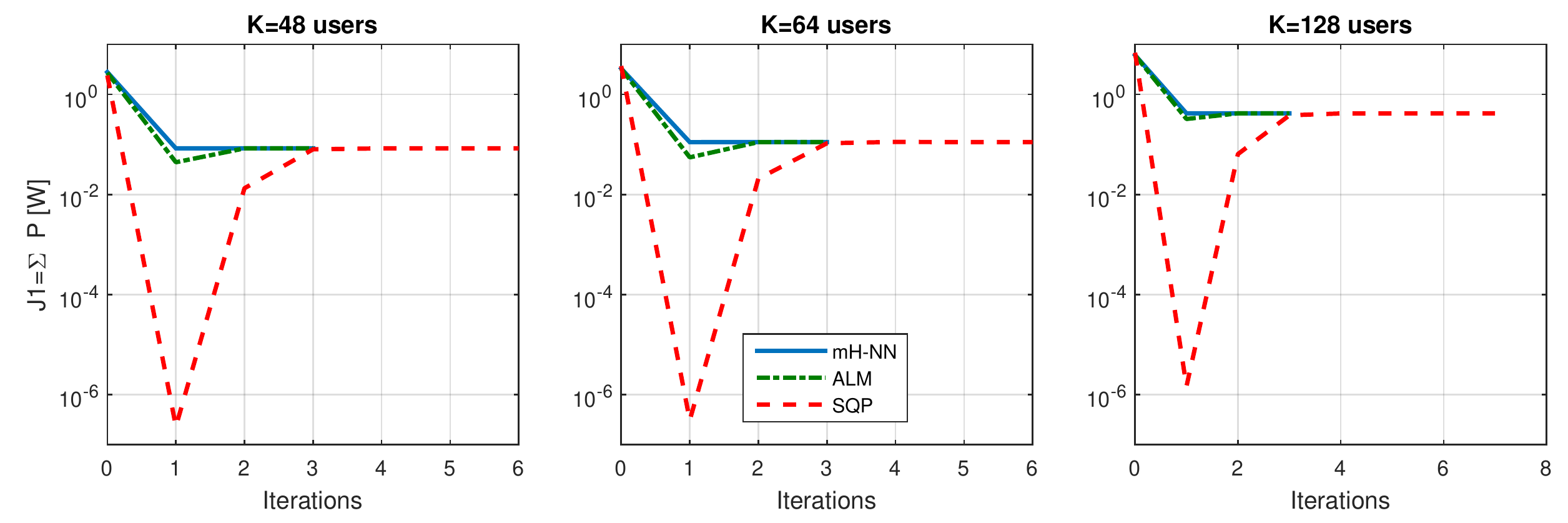}
(c) $\Gamma^*= 22$dB
\caption{Sum-power levels for $K=48, \ 64$ and $128$ users and different $\Gamma^*\in[17, \,\, 20 ,\,\, 22]$dB.}
\label{K48,64,128_17_20_22db}
\end{figure}

\begin{figure}[!htbp]
\centering
\small
\includegraphics[width=17.5cm]{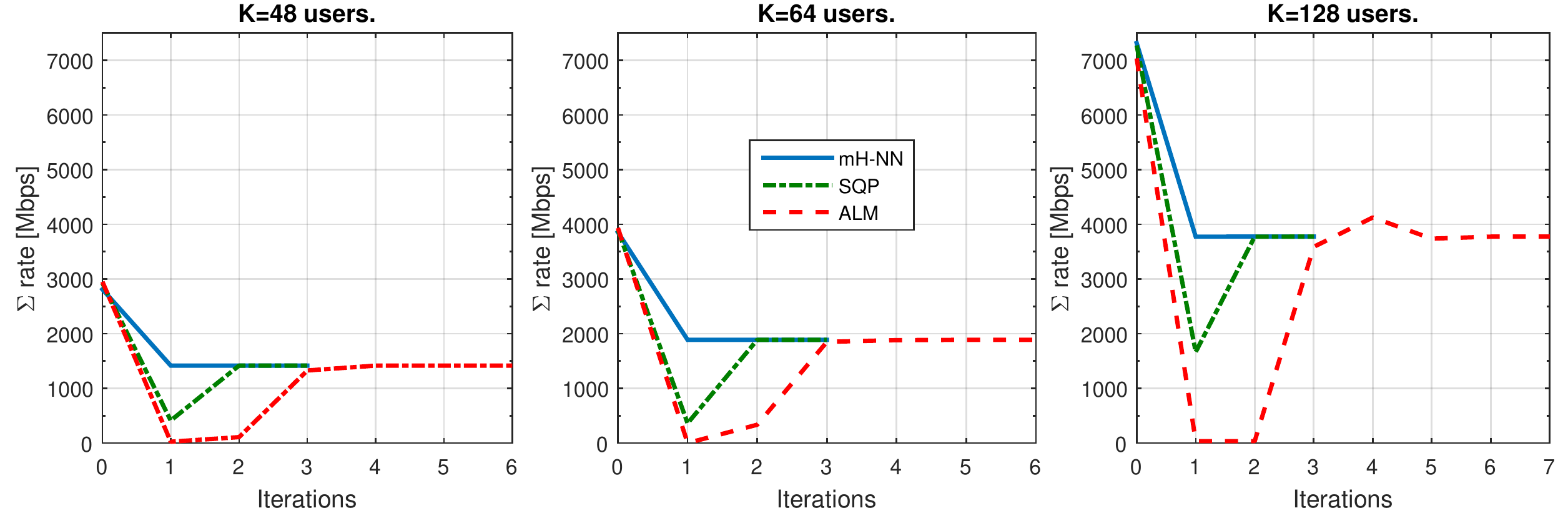}
(a) $\Gamma^*= 17$dB\\
\vspace{4mm}
\includegraphics[width=17.5cm]{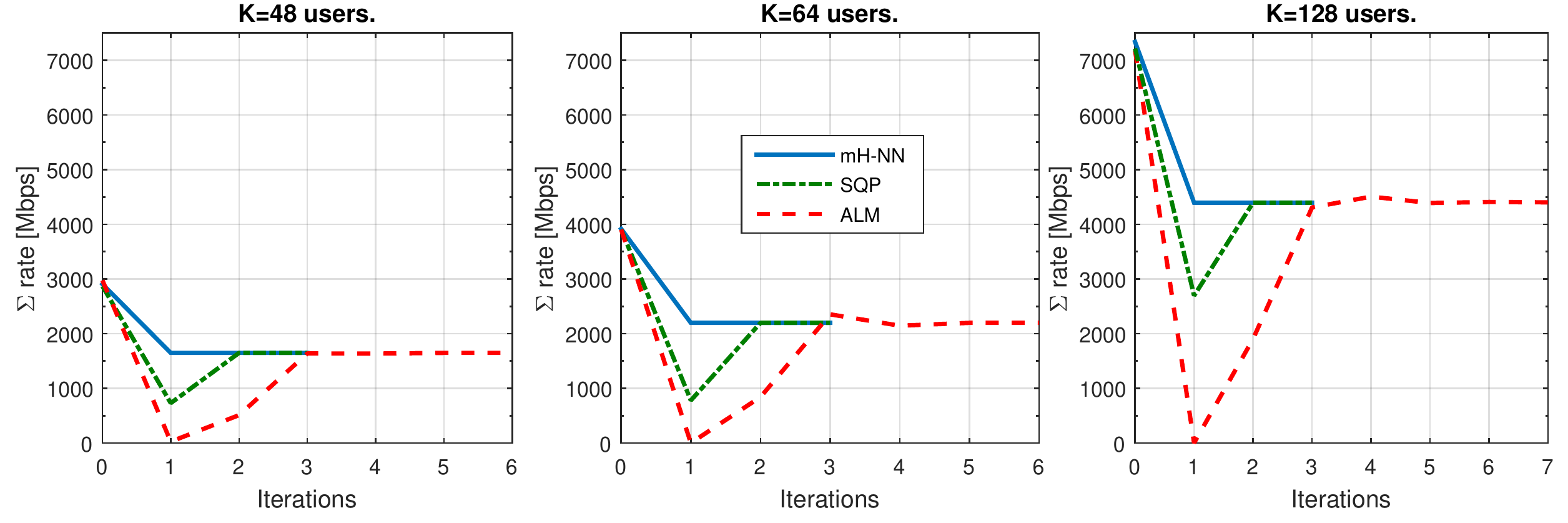}
(b) $\Gamma^*= 20$dB\\
\vspace{4mm}
\includegraphics[width=17.5cm]{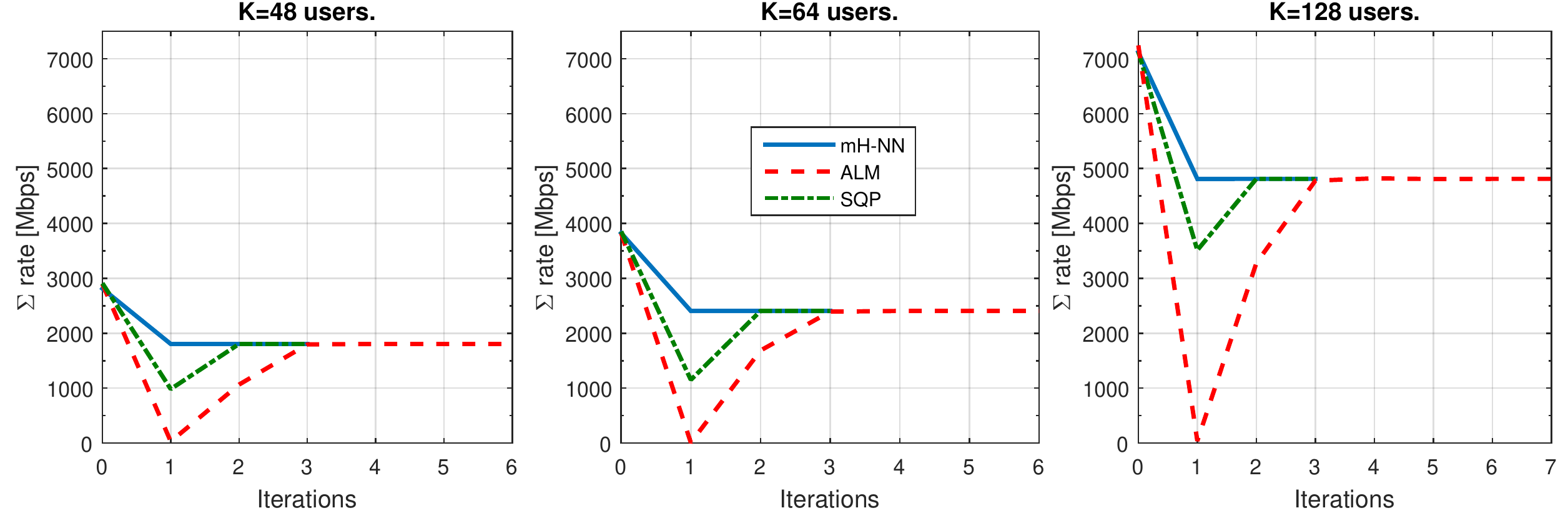}
(c) $\Gamma^*= 22$dB
\caption{Sum-rate allocation across the iterations in for $K = 48; 64$ and $128$
users (Scenario B) with different target SINR: a) $\Gamma^* =17$ dB; b) $\Gamma^* =20$ dB, and c) $\Gamma^* =22$ dB.}
\label{K48,64,128_17_20_22sumrate}
\end{figure}

\newpage
\subsection{Complexity Analysis}
The quality of the solutions achieved by three algorithm is evaluated through the NMSE metric presented in Tables \ref{tabelares} and  \ref{tabelaresB1-B3}. 
Figure \ref{MSE} also shows the analysis of NMSE evolution as a function of the number of interactions for Scenario A considering systems with $K =8$, $16$ and $32$ users. One can observe that the best MSE levels occur for SQP and mH-NN. This behavior was repeated in Scenario B.

\begin{figure}[!htbp]
\begin{center}
\includegraphics[trim={.1mm 1mm 1mm 1mm},clip,width=1\textwidth]{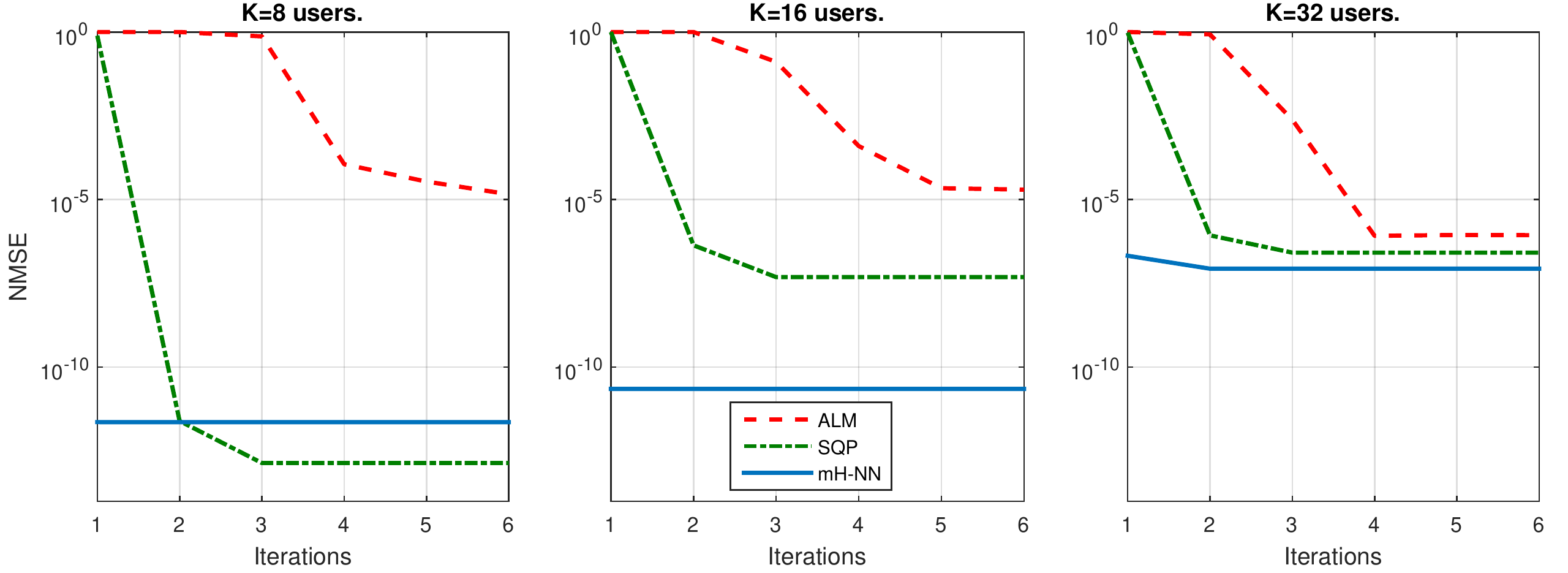}
\caption{NMSE evolution for the methods addressed in relation to the power vector $\p^*$ for $U=[8, 16, 32]$ users.} 
\end{center}
\end{figure}

Fig. \ref{tempo_ex} put in perspective the computational time consumption for the modified Hopfield network algorithm compared to the SQP for all $K=\{8, 16, 32, 48, 64, 128\}$ users configurations, highlighting its superiority over ALM for the all scenarios evaluated. We can observe that the implementation of the mH-NN Algorithm \ref{Hopfield} is simple when compared to the ALM implementation, as well as the SQP method. The augmented Lagrangian method had lower performance for all users scenarios and it is worth mentioning that, based on results found in the literature, the use of a more sophisticated implementation of ALM, for instance, considering an ALM-based solver like ALGENCAN \cite{37} would reduce the time burden processing, becomes close to those obtained by the mH-NN and SQP methods.
\begin{figure}[!htbp]
\centering
\includegraphics[width=12cm]{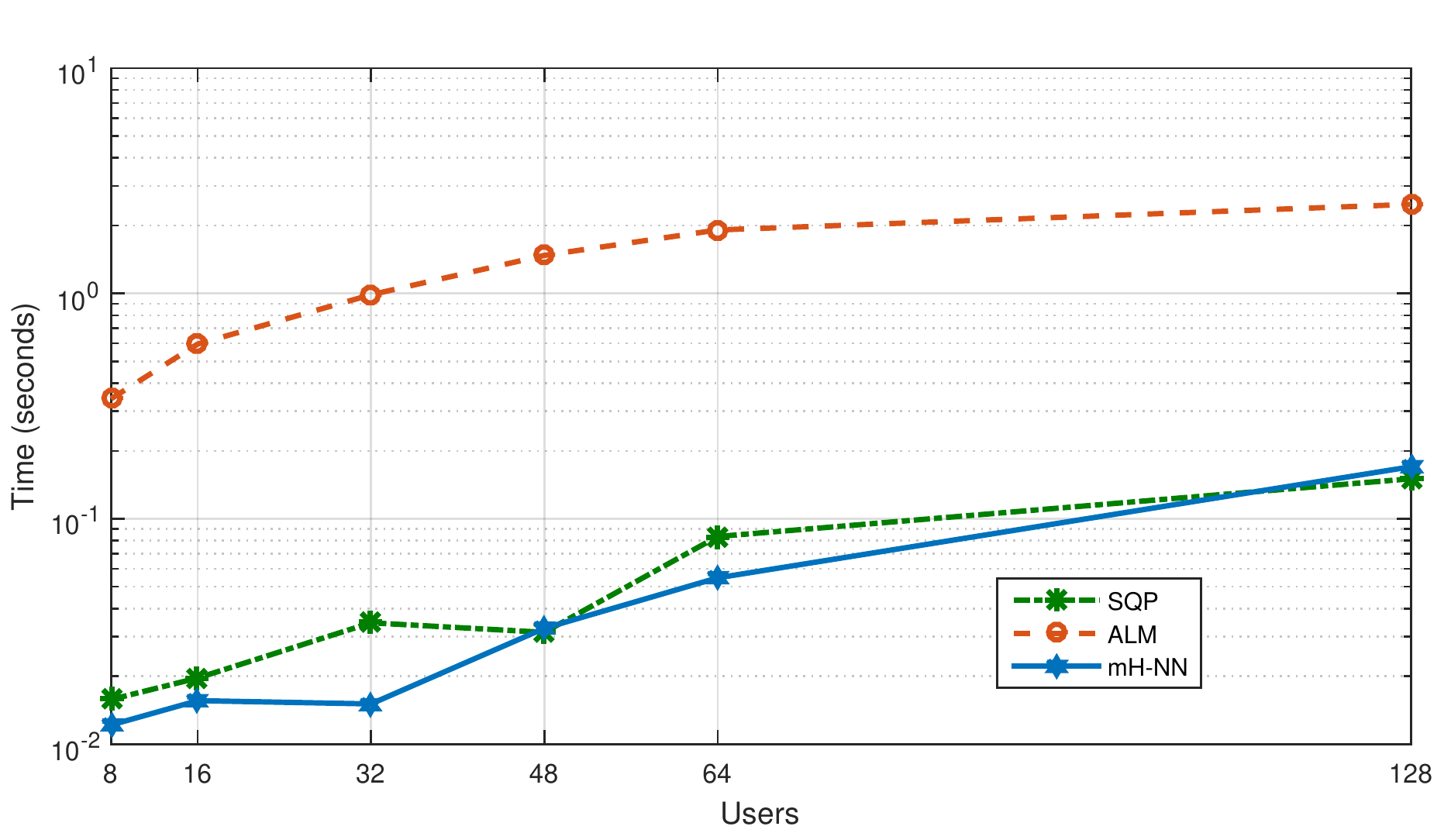}
\vspace{-3mm}
\caption{Execution time for the three methods considering $K=\{8, 16, 32, 64, 128\} $ users and QoS level $20$ dB.}
\label{tempo_ex}
\end{figure}

To bring more insight on the three algorithms' complexity, Table \ref{tab:robustez} summarizes the algorithm robustness, measured as defined in (\ref{R}), considering Scenario B with QoS rate $22$dB for $100$ realizations. The robustness obtained demonstrated that the SQP and mH-NN methods result in full convergence success under the considered high loading system scenario, while the ALM was able to attain convergence to acceptable solutions in the majority of the realizations. There are some cases where the ALM is not able to reach full convergence, meaning that the method does not attain a feasibility level of $\mathcal{F} = 10^{-4}$, according to the stipulated feasibility criterion. Moreover, the stopping criterion was maintained in this context, i.e., given by eq. (\ref{criterio_parada}), while the maximum number of external iterations $\mathcal{I}=10$ was stipulated.

\begin{table}[!htbp]
\centering
\caption{Robustness of the ALM, SQP and mH-NN algorithms for the power allocation problem in eq. (\ref{probe}) over $\mathcal{T}=100$ realizations, considering QoS level of 22dB.}
\label{tab:robustez}
{{\renewcommand{\arraystretch}{1.2}%
\resizebox{0.35\textwidth}{!}{%
\begin{tabular}{|c|c|c|c|}
\hline \bf Users & {$\mathcal{R}$-ALM} & {$\mathcal{R}$-SQP} & {$\mathcal{R}$-mH-NN} \\ \hline \hline
\hline 48           &         100$\%$     &          100$\% $   &        100$\%$      \\
\hline 64           &          98$\%$     &          100$\% $   &        100$\%$      \\
\hline 128          &          93$\%$     &          100$\%$    &        100$\%$     \\
\hline
\end{tabular}
}}}
\end{table}

In the equilibrium, the system power allocation solution given by eq. \eqref{tarhuni}, the matrix $-\bf \Gamma^* \H$ may have entries close to zero, since this is obtained through the gain matrix and target CIR, depending on the scale used. When we perform the matrix sum $(\I-\bf \Gamma^* \H)$, small loss of information may occur. In this sum, one can lose information due to the order of magnitude of the entries of the matrix $(-\bf \Gamma^* \H)$. Hence, in the resolution of the system (\ref{tarhuni}), due to the propagation errors phenomenon, especially considering network with large number of users, namely $K = 48, \, 64$ and $128$ users, the Tarhuni solution \eqref{tarhuni} fails to achieve feasibility levels as obtained by mH-NN and SQP as depicted in Table \ref{tabelaresB4}.

\begin{center}
\begin {table}[!htbp]
\small
\centering
\caption {Class C -- Sum-power and attainable feasibility from eq. (\ref{tarhuni}) for  increasing number of optical nodes.}
 \label{tabelaresB4}
{{\renewcommand{\arraystretch}{1.2}%
\resizebox{0.8\textwidth}{!}{%
\begin {tabular}{|c|c|c|c|c|c|}
\hline     \bf \# Users & $J_1(\textbf{p})$ [mW]  & \bf Feasibility, $\mathcal{F}$ & \bf Tarhuni $\Sigma$-Power [mW] & \bf Tarhuni  Feasibility $\mathcal{F}$ \\ \hline
\hline           48 &            83.42            &  1.20122e-17                  &          83.38             &       2.51414e-04                     \\
\hline           64 &            111.16           &  1.55158e-17                  &          111.01            &       3.901331e-04                   \\
\hline           128 &           417.61           &  1.55134e-17                  &           417.36           &       6.007099e-04                    \\
\hline
\end {tabular}
}}}
\end {table}
\end{center}

\subsubsection{Dynamical Performance Analysis}
The power variations in the network are related to the linear and non-linear effects associated with the optical fiber, as well as to the coupling effects of channel power, which are influenced by the network topology, traffic variation and physics of optical amplifiers, as well as dynamic addition and removal of channels. In addition, there are the effects of the unpredictability of time-varying penalties, such as polarization effects \cite{31}.

In this subsection, we extend the PC algorithms robustness analysis by analyzing in a more challenging power allocation scenario considered a larger number of users and different QoS requirements. Also, on fly modifications were introduced in the operational configurations of the optical network which parameters are presented in the Table \ref{tab:param3}, but now taking into account the dynamic addition of channels. For this analysis, after convergence of the three methods considered $32$ users and QoS level of $22$dB, the number of users was increased by $300\%$ reaching 128 active users in the OCDMA network.

\begin{figure}[!htbp]
\begin{center}
\includegraphics[trim={4mm 5mm 10mm 1mm},clip,width=.32\textwidth]{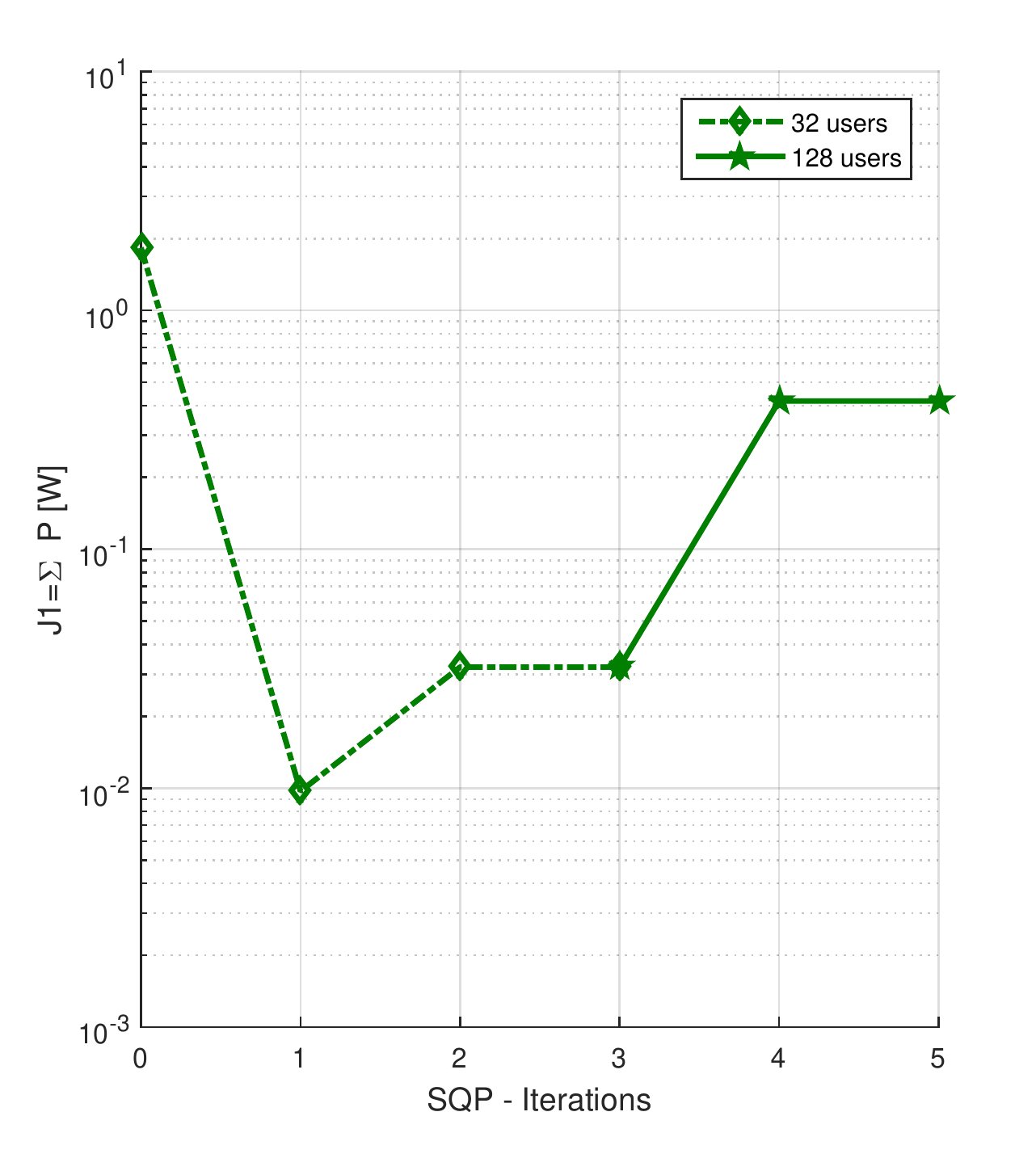}
\includegraphics[trim={3mm 4mm 11mm 1mm},clip,width=.32\textwidth]{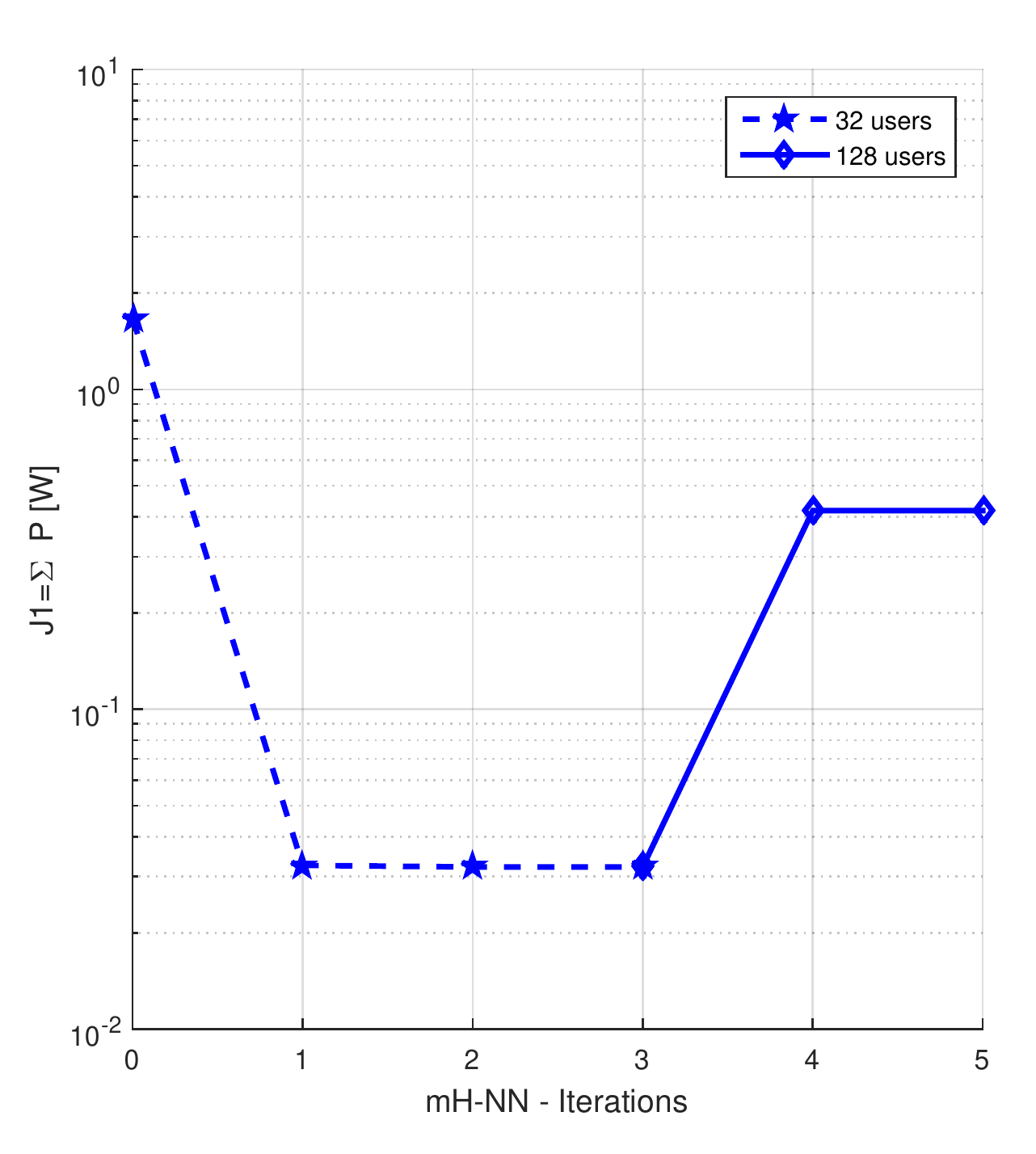}
\includegraphics[trim={3mm 5mm 11mm 1mm},clip,width=.32\textwidth]{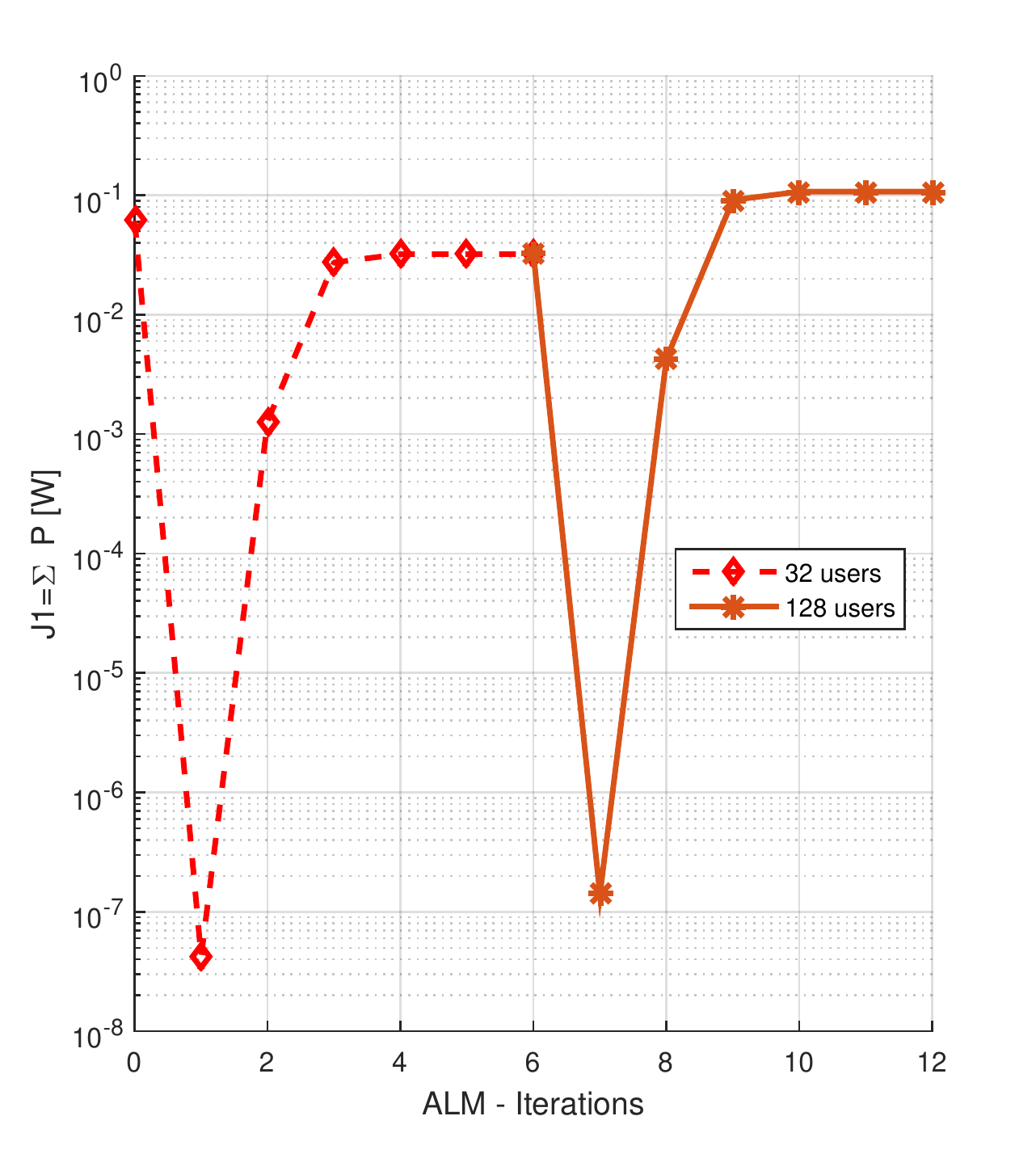}
{\small a) SQP \hspace{4.2cm} b) Hopfield  \hspace{4.2cm}  c) ALM}
\vspace{-1mm}
\caption{Behavior of the a) SQP and b) Hopfield and c) ALM power allocation methods under the on fly increasing number of users in $300\%$.} 
\label{onfly_sqpH_alm}
\end{center}
\end{figure}

In Fig. \ref{onfly_sqpH_alm}, it can be observed that after the restart with 128 users, the SQP, mH-NN and ALM methods respectively consumed 2, 2 and 6 iterations to reach full convergence in the new 128 user's power allocation equilibrium. Besides, the ALM  behavior reveals that as it was taking into account a restart of the method not taking advantage of the previous power allocation solution for 32 users, thus consuming 6 a high number of iterations to reach the convergence. On the other hand SQP and Hopfield needed just only 2 iterations to fully achieve the new equilibrium.

In order to evaluate the capability of power allocation algorithms  to re-establish the power equilibrium after a strong perturbation,  it was considered an optical power perturbation in the power in the $i$th user, modeled as:
\begin{equation}\label{eq:perturb}
p_i[n] = |\alpha^n \cdot {\rm sin} \left(1.5\pi \cdot n\right)| + p_i^{\circ}, \qquad n\geq0
 \end{equation}
where $\alpha=0.65$, $p_i^{\circ}$ is the nominal transmitted power for the $i$th lightpath and $n$ represents the current iteration. For illustration purpose, Fig. \ref{fc_pertubation} depicts the perturbation function (\ref{eq:perturb}) considering $p_i^{\circ} = 0.1$ W. It can be seen that the effect of the disturbance tends to disappear as we increase the value of $n$. This perturbation was considered in the first half of the iterations, i.e., $2\leq n \leq 7$, that would be spent by the three methods to obtain the convergence, for $K=128 $ users and QoS of $22$dB. Notice that under external perturbation regime, $\mathcal{I}=15$ iterations has been adopted. For the SQP and Hopfield methods, the perturbation was included starting from the second iteration, while for the ALM method it was included  starting from the third iteration.

\begin{figure}[!htbp]
\begin{center}
\includegraphics[width=10cm]{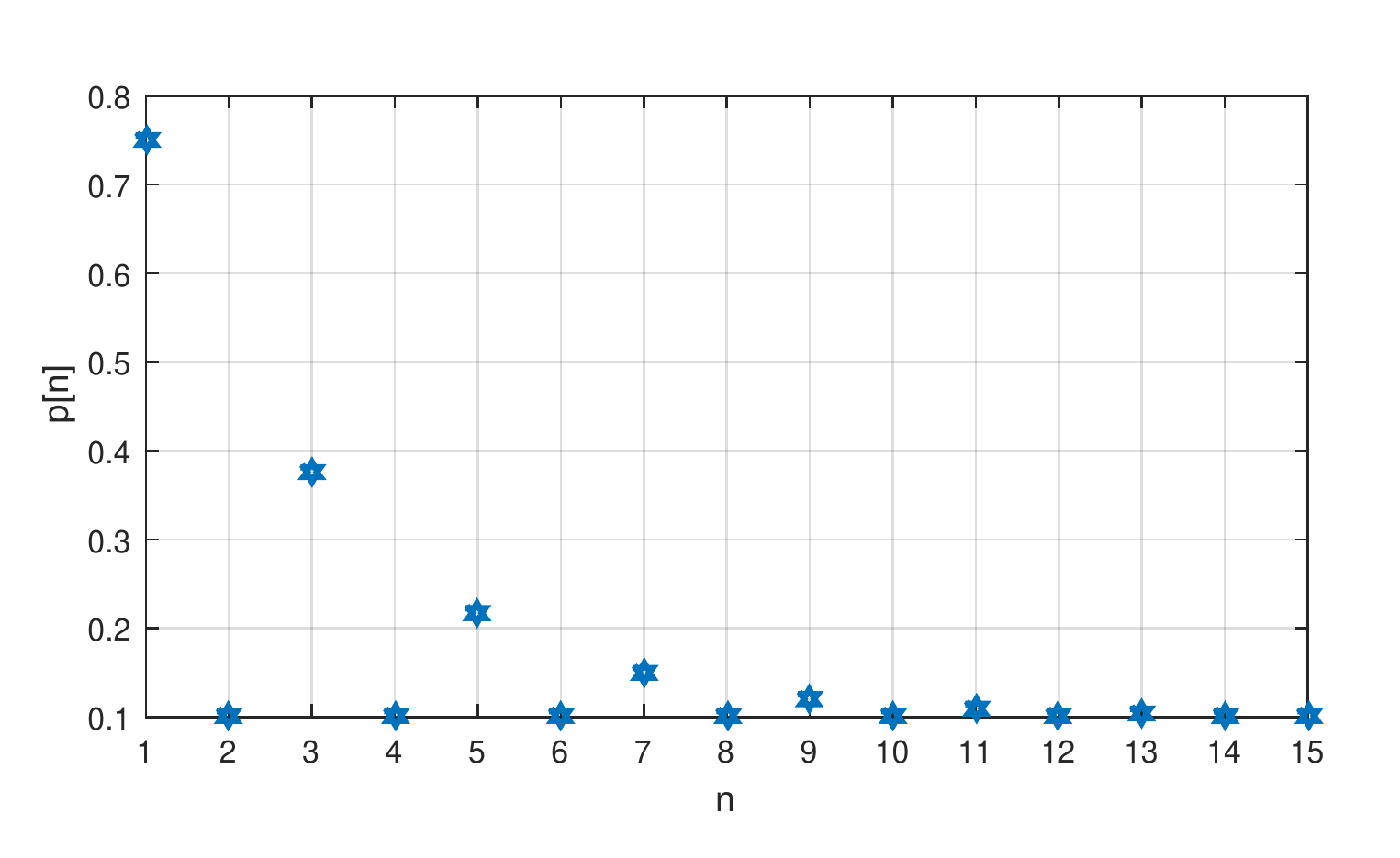}
\vspace{-3mm}
\caption{Perturbation function $p_i[n]$ illustration $\times$ number of iteration $n$ with $p_i^{\circ} = 0.1$ W.}
\label{fc_pertubation}
\end{center}
\end{figure}

The power control response to the disturbance is evaluated along the 15 iterations, as depicted in Figure \ref{perturbacao}. One can notice that the three OCDMA power allocation methods are able to recover to the perturbation introduced by eq. \eqref{eq:perturb}, re-establishing the convergence equilibrium to the optimum power allocation $\bf p^*$, but with perturbation effects on the convergence process, mainly noted at the beginning iteration instants where perturbation is inserted. Indeed, modified Hopfield-based method meets the convergence criterion very early, after 10 iterations, while ALM attains convergence after 14 iterations, and SQP after 12 iterations. Thus, we can see that Hopfield needed 1 iteration to reach equilibrium and 3 more to converge following the $\mathcal{F}$ and $\xi$ value criteria defined in Table \ref{tab:param3}.

\begin{figure}[!htbp]
\centering
\includegraphics[width=18cm]{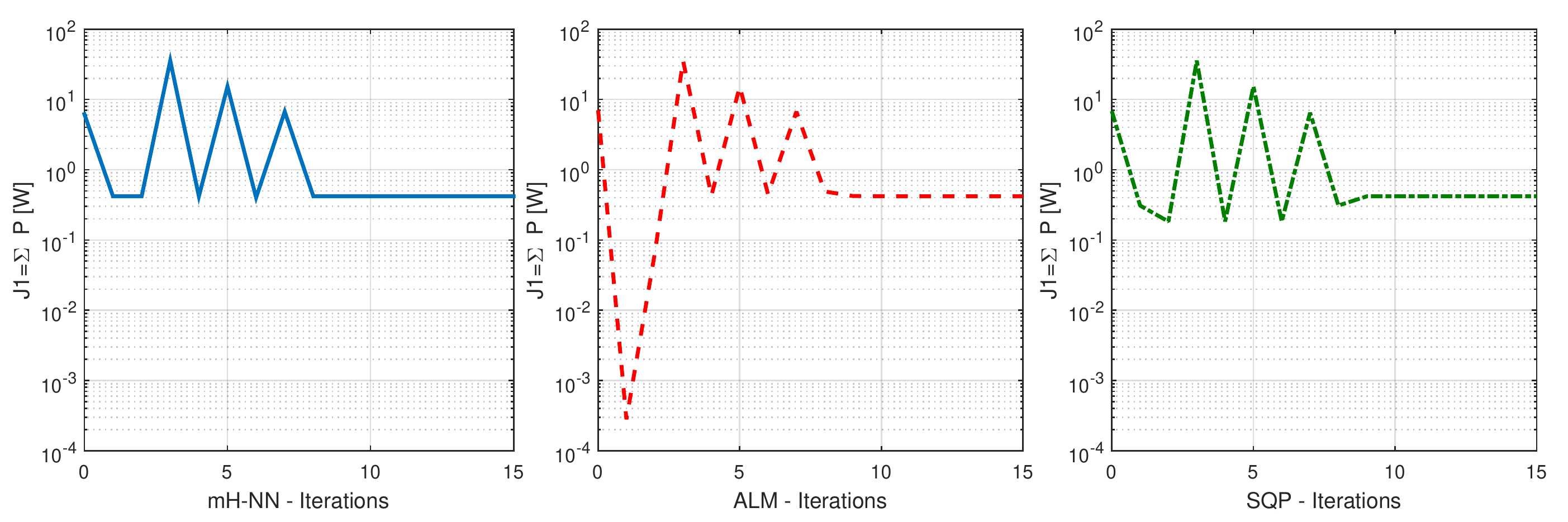}
\vspace{-8mm}
\caption{Representative power allocation convergence per ONU considering a perturbation of eq. \eqref{fc_pertubation} inserted at $2\leq n\leq 7$ iteration for the Hopfield, SQP and ALM methods.}
\label{perturbacao}
\end{figure}

\section{Conclusion}\label{sec:Concl}
In this work it was demonstrated that the methodologies studied are adequate for the problem of minimum power allocation in OCDMA optical networks operating under different system loading scenarios. We highlight that the proposed modified Hopfield network proved to be an effective alternative to solve the power allocation problem in OCDMA networks when compared to the classic programming methods due to its low computational cost, while its simplicity of implementation does not require previous training.

It is also worth noting that the conventional Hopfield network usually consumes many iterations to achieve the convergence, motivating us to use the direction of the gradient to optimize the objective function in Step III of the Algorithm \ref{Hopfield}. Moreover, in the problem (\ref{probe}) the gradient is the only direction of descent since the function is linear which justifies the fact that the method has consumed few iterations to obtain convergence.

While SQP method takes advantage of the simplicity of the objective function for the construction of simpler subproblems, on the other hand the ALM minimizes the Lagrangian function in the box $p_{\min}\leq \p \leq p_{\max}$, which is not simplified due to the characteristics of the objective function. In this way, among the representative NLP methods evaluated, one can always expect a better performance of the SQP method with respect to the ALM in solving the power allocation problem in OCDMA networks.

\section*{Acknowledgment}
This work was supported in part by the National Council for Scientific and Technological Development (CNPq) of Brazil underGrant 304066/2015-0, by Universidade Tecnológica Federal do Paraná -- UTFPR, Campus Cornélio Procópio, and in part by State University of Londrina -- Paraná State Government (UEL).






\end{document}